\documentclass[aip,jcp,reprint,amsmath,amssymb,floatfix]{revtex4-2}
\usepackage[T1]{fontenc}
\usepackage[utf8]{inputenc}
\usepackage{mathtools,amsthm,bm,booktabs,dcolumn,graphicx,microtype,xurl}
\usepackage[hidelinks]{hyperref}
\newtheorem{proposition}{Proposition}
\newcommand{\Nop}{\hat N_\Omega}
\newcommand{\numdhbebChargePqg}{+0.41}
\newcommand{\numdhbebChargePqgReg}{+0.06}
\newcommand{\numdhbebChargeTt}{+0.38}
\newcommand{\numdhbebChargeTtReg}{0.00}
\newcommand{\numdhbebCutOnset}{2.75}
\newcommand{\numdhbebNR}{26}
\newcommand{\numdhbebOrbOnset}{3.75}

\newcommand{\numdhbebPQGRinf}{$-$43.33}

\newcommand{\numdhbebPQGcutRFourFiveZero}{$-$3.94}
\newcommand{\numdhbebPQGcutRFiveZeroZero}{$-$3.19}

\newcommand{\numdhbebPQGcutRNineZeroZero}{$-$2.64}

\newcommand{\numdhbebPQGorbRFiveZeroZero}{$-$9.80}

\newcommand{\numdhbebPQGorbRNineZeroZero}{$-$2.64}

\newcommand{\numdhbebRefDevFrom}{3.75}
\newcommand{\numdhbebRefDevMax}{0.18}
\newcommand{\numdhbebRefDevMid}{3.9}
\newcommand{\numdhbebRefDevTo}{6.00}

\newcommand{\numdhbebRefPlusFourFive}{$-$18.73}
\newcommand{\numdhbebRefPlusNine}{$-$0.37}
\newcommand{\numdhbebRmax}{9.00}
\newcommand{\numdhbebRmin}{1.25}

\newcommand{\numdhbebTOneTwocutRNineZeroZero}{$-$0.007}

\newcommand{\numdhbebTtRegMax}{1.01}

\newcommand{\numdhlifPqgOrbMax}{0.06}

\newcommand{\numdhnoCutOnset}{2.00}
\newcommand{\numdhnoNR}{24}
\newcommand{\numdhnoOrbOnset}{2.25}

\newcommand{\numdhnoPQGRinf}{$-$144.61}

\newcommand{\numdhnoPQGcutRThreeZeroZero}{$-$49.33}

\newcommand{\numdhnoPQGcutRSixZeroZero}{$-$42.81}

\newcommand{\numdhnoPQGorbRThreeZeroZero}{$-$60.21}

\newcommand{\numdhnoPQGorbRSixZeroZero}{$-$42.64}

\newcommand{\numdhnoRmax}{8.00}
\newcommand{\numdhnoRmin}{0.90}

\newcommand{\numdhnoTtRegMax}{17.33}
\newcommand{\numdhlifCutOnset}{3.0}
\newcommand{\numdhlifGapMin}{20}
\newcommand{\numdhlifIonicAbove}{131}
\newcommand{\numdhlifNLiEq}{2.19}
\newcommand{\numdhlifNLiLong}{2.98}
\newcommand{\numdhlifNR}{18}
\newcommand{\numdhlifPqgCutMax}{0.08}

\newcommand{\numdhlifPqgErrMax}{24.9}
\newcommand{\numdhlifRCross}{4.5}
\newcommand{\numdhlifRmax}{10.0}
\newcommand{\numdhlifRmin}{1.2}
\newcommand{\numdhlifTtCutMax}{0.01}
\newcommand{\numdhlifTtErrMax}{1.3}
\newcommand{\numRdGuMax}{5.8\times10^{-4}}
\newcommand{\numRdGvMax}{1.3\times10^{-3}}
\newcommand{\numRdLambdaMax}{11.8}
\newcommand{\numRdLambdaMin}{10.0}
\newcommand{\numRdLztwoMax}{1.5\times10^{-7}}
\newcommand{\numRdNsol}{232}

\newcommand{\numRdPmCNmSixErr}{$-$76.33}
\newcommand{\numRdPmCNmSixPartIErr}{$-$76.42}
\newcommand{\numRdPmCandMax}{22077}
\newcommand{\numRdPmCandMin}{20841}
\newcommand{\numRdPmMaxDiff}{0.08}

\newcommand{\numRdPmNOpSixErr}{$-$78.17}
\newcommand{\numRdPmNOpSixPartIErr}{$-$78.19}

\newcommand{\numRdPmNtwopFourErr}{$-$137.98}
\newcommand{\numRdPmNtwopFourPartIErr}{$-$137.99}
\newcommand{\numRdSmSpMax}{4.6\times10^{-7}}

\newcommand{\numRdStatusOpt}{66}
\newcommand{\numRdZAllGapMax}{0.02}
\newcommand{\numRdZBcpRepro}{0.05}

\newcommand{\numRdZCNmFourMaxErrInner}{0.63}

\newcommand{\numRdZCNmThreeBcpErr}{$-$5.33}

\newcommand{\numRdZCNmThreeOptErr}{$-$4.79}
\newcommand{\numRdZCNmThreeOptZ}{0.44}

\newcommand{\numRdZCNmTwoFiveBcpErr}{$-$21.98}

\newcommand{\numRdZCNmTwoFiveOptErr}{$-$21.94}
\newcommand{\numRdZCNmTwoFiveOptZ}{0.48}

\newcommand{\numRdZNOpFourMaxErrInner}{0.15}

\newcommand{\numRdZNOpThreeBcp}{0.55}
\newcommand{\numRdZNOpThreeBcpErr}{$-$3.52}
\newcommand{\numRdZNOpThreeBestErr}{$-$1.05}
\newcommand{\numRdZNOpThreeBestZ}{0.45}

\newcommand{\numRdZNOpThreeOptErr}{$-$0.99}
\newcommand{\numRdZNOpThreeOptZ}{0.43}

\newcommand{\numRdZNOpTwoFiveBcp}{0.56}
\newcommand{\numRdZNOpTwoFiveBcpErr}{$-$15.77}
\newcommand{\numRdZNOpTwoFiveBestErr}{$-$9.15}
\newcommand{\numRdZNOpTwoFiveBestZ}{0.45}

\newcommand{\numRdZNOpTwoFiveOptErr}{$-$8.93}
\newcommand{\numRdZNOpTwoFiveOptZ}{0.47}

\newcommand{\numBeOpFciChargeFour}{1.00}
\newcommand{\numBeOpHfChargeFive}{1.00}
\newcommand{\numBeOpHfChargeFour}{0.57}
\newcommand{\numBeOpOrbFive}{0.2}
\newcommand{\numBeOpOrbFour}{31.6}
\newcommand{\numBeOpRegFive}{1.6}
\newcommand{\numBeOpRegFour}{1.3}

\newcommand{\numCNmPqg}{80.7}

\newcommand{\numCOdipFci}{0.00}

\newcommand{\numCOdipPqgLz}{0.47}

\newcommand{\numCOdipPqgReg}{0.02}

\newcommand{\numCOdipTtLz}{0.63}

\newcommand{\numCOdipTtRegLz}{0.00}

\newcommand{\numCandMax}{1160}
\newcommand{\numCandMin}{231}
\newcommand{\numEigExcess}{$6\times10^{-10}$}
\newcommand{\numEigLBound}{$4\times10^{-7}$}

\newcommand{\numFinalOrbViolMaxPqg}{$9.9\times10^{-7}$}
\newcommand{\numFinalOrbViolMaxTt}{$3.5\times10^{-5}$}

\newcommand{\numFinalRegViolMaxPqg}{$7.0\times10^{-8}$}
\newcommand{\numFinalRegViolMaxTt}{$7.8\times10^{-8}$}

\newcommand{\numGapMax}{0.28}
\newcommand{\numGapMedian}{0.002}
\newcommand{\numGridDS}{$2\times10^{-4}$}
\newcommand{\numGridDq}{$2\times10^{-3}$}

\newcommand{\numGridEnergy}{0.05}
\newcommand{\numGridEnergyAbs}{0.002}

\newcommand{\numLzCOSix}{0.98}
\newcommand{\numLzCOSixShift}{0.02}
\newcommand{\numLzMaxCharge}{0.04}
\newcommand{\numLzMaxShift}{7.2}
\newcommand{\numLzMinShift}{$-$0.02}
\newcommand{\numLzNOpFour}{0.61}
\newcommand{\numLzNOpFourShift}{0.27}
\newcommand{\numLzNabove}{25}
\newcommand{\numLzPoints}{137}
\newcommand{\numLzTtImposed}{5}
\newcommand{\numLzTtMaxShift}{0.3}
\newcommand{\numLzTtOrbShift}{5.3}
\newcommand{\numLzTtPoints}{5}
\newcommand{\numMCOPqgRegMu}{6.001}
\newcommand{\numMCOPqgRegSA}{2.00}
\newcommand{\numMCOPqgRegSB}{2.00}
\newcommand{\numMCOpFciSA}{0.75}

\newcommand{\numMCOpPqgRegMu}{5.052}
\newcommand{\numMCOpPqgRegVar}{0.050}
\newcommand{\numMCOpTtRegMu}{5.000}
\newcommand{\numMCOpTtRegSA}{0.92}
\newcommand{\numMCOpTtRegSB}{2.00}
\newcommand{\numMaxRegionCuts}{3}

\newcommand{\numNOpPqg}{108.9}

\newcommand{\numNtwoPlusPqg}{138.0}

\newcommand{\numNtwoPlusTtReg}{1.6}

\newcommand{\numPartMaxDiff}{0.2}
\newcommand{\numPartPoints}{4}
\newcommand{\numRegBound}{0.02}
\newcommand{\numSCCOPqgReg}{5.9}
\newcommand{\numSCCOTtReg}{0.01}
\newcommand{\numSCCOpPqgReg}{19.5}
\newcommand{\numSCCOpTtReg}{5.0}
\newcommand{\numSCFragMax}{0.013}
\newcommand{\numSCNOpPqg}{130.6}
\newcommand{\numSCNoCutAdditive}{3}
\newcommand{\numSCNtwoPPqgReg}{21.5}
\newcommand{\numSCNtwoPTtReg}{1.4}

\newcommand{\numSCOtherPqgReg}{0.2}
\newcommand{\numSCOtherTtReg}{0.01}
\newcommand{\numSCPoints}{10}
\newcommand{\numSCRestored}{7}

\newcommand{\numSCZeroPqg}{4}
\newcommand{\numSCZeroTt}{7}
\newcommand{\numSNOpFourFci}{3.75}
\newcommand{\numSNOpFourPqg}{2.96}
\newcommand{\numSNOpFourPqgReg}{3.75}
\newcommand{\numSNtwoPlusFourFci}{2.87}
\newcommand{\numSNtwoPlusFourPqg}{2.93}
\newcommand{\numSNtwoPlusFourPqgReg}{3.07}
\newcommand{\numSNtwoPlusFourTtReg}{2.90}

\newcommand{\numTNOpLowdinOrbL}{$-$0.181}
\newcommand{\numTNOpPmFciVar}{0.046}
\newcommand{\numTNOpPmOrbMu}{6.68}
\newcommand{\numTNOpPmOrbVar}{0.218}
\newcommand{\numTNOpPmPqgMu}{6.58}
\newcommand{\numTNOpPmPqgVar}{0.040}
\begin{document}
\title{Integer electron-count cuts in variational second-order reduced-density-matrix theory.
I.~Molecular dissociation with fixed real-space and orbital partitions}
\author{Maho Nakata}
\email{maho@riken.jp}
\affiliation{RIKEN Cluster for Pioneering Research, Wako, Saitama 351-0198, Japan}
\date{\today}
\begin{abstract}\label{rev:abstract}
Van Aggelen \emph{et al.} [Phys.\ Chem.\ Chem.\ Phys.\ \textbf{11}, 5558 (2009)]
showed that variational second-order reduced-density-matrix (2-RDM)
calculations with the standard two-index conditions dissociate diatomic
molecules into fractionally charged atoms with energies far below the exact
limit, a drastic violation of size consistency; the proposed subsystem
constraints require auxiliary calculations with the physical fragment
Hamiltonians. We show that two long-known ingredients, combined, remove most of this
failure without such calculations: the requirement that the number of electrons in a part of the
molecule be an integer, imposed as linear conditions on the 2-RDM, and a
chemically motivated partition of the molecule into atomic regions bounded at
the bond critical point of the Hartree--Fock density. In a minimal basis,
compared with full configuration interaction, a single such condition reduces
the energy errors of NO$^+$ and CN$^-$ at 6~\AA{} from \numNOpPqg{} and
\numCNmPqg~m$E_\mathrm{h}$ to less than \numRegBound~m$E_\mathrm{h}$. In a
split-valence basis, combined with the three-index conditions, they keep BeB$^+$
within \numdhbebTtRegMax~m$E_\mathrm{h}$ of FCI at all bond lengths.
In N$_2^+$, whose fragments have the correct mean population but too small a
fluctuation, the region and three-index conditions act complementarily,
reducing the error at 4~\AA{} from \numNtwoPlusPqg{} to
\numNtwoPlusTtReg~m$E_\mathrm{h}$. At the separated-atom limit, the region
cuts make the $P$, $Q$, $G$ energy additive over the fragments to within
\numSCOtherPqgReg~m$E_\mathrm{h}$ in \numSCRestored{} of \numSCPoints{}
systems, \numSCNoCutAdditive{} already nearly additive without cuts; adding the $T_1$ and $T_2$ conditions also makes CO nearly additive,
while a nonadditive error remains for N$_2^+$ and CO$^+$.
\end{abstract}
\maketitle

\section{Introduction}

Variational optimization of the second-order reduced density matrix (2-RDM)
under necessary $N$-representability conditions\cite{Coleman1963} gives lower
bounds to the ground-state energy.\cite{Garrod1964,RosinaGarrod1975,GarrodFusco1976,Erdahl1978}
Semidefinite programming made it a practical method,\cite{Nakata2001,Zhao2004,Mazziotti2004}
first-order algorithms extended it to large systems,\cite{Mazziotti2011} and
Ref.~\citenum{MazziottiBook2007} reviews the field. The standard conditions
are, however, known to fail for simple model Hamiltonians\cite{Nakata2006} and
for molecular dissociation. Van Aggelen, Bultinck, Verstichel, Van Neck, Ayers,
and Cooper showed that with the two-index conditions $P$, $Q$, and $G$ the
method dissociates heteronuclear diatomic molecules into fractionally charged
atoms with too low an energy,\cite{VanAggelen2009,VanAggelen2010} a counterpart
of the fractional-charge and fractional-spin errors of approximate density
functionals.\cite{Perdew1982,Cohen2008} The three-index conditions $T_1$ and
$T_2$ do not remove this failure, and a recent tutorial review still lists it
as a difficulty of the method.\cite{DePrince2024} Subsystem constraints, which
require the energy of an atomic subspace not to fall below the convex
interpolation of separately computed fragment energies, restore the correct
dissociation limit of diatomic molecules,\cite{Verstichel2010,VanAggelen2011}
but they require auxiliary calculations with the physical Hamiltonian of every
fragment and for every charge, and the constraints needed for polyatomic
molecules depend on the geometry.\cite{VanAggelen2011} An earlier analysis of
size consistency showed that a 2-RDM is size-consistent if, among other
conditions, the reduced density matrices of the fragments are eigenstates of
the fragment electron numbers;\cite{Nakata2011} imposing this condition
requires the integer charges of the fragments to be known in advance, which is
part of the problem to be solved.

In this paper we use a weaker, Hamiltonian-independent consequence of the same
physical requirement: the number of electrons found in a fixed region is an
integer-valued random variable. For every integer $k$ this gives a linear
inequality on the first- and second-order RDMs that does not prescribe which
integer is realized. Inequalities of this type are known: for real-space
regions they are members of the family of necessary conditions for pair
distribution functions given by Ayers and Davidson,\cite{AyersDavidson2006}
and for sums of orbital occupation numbers they belong to the diagonal
conditions\cite{WeinholdWilson1967,McRaeDavidson1972,Davidson2003} reviewed in Ref.~\citenum{AyersDavidson2007}.
The bound on the variance of an integer count used below goes back to
Yamada,\cite{Yamada1961} and necessary conditions on the electron pair density
were also derived by Davidson.\cite{Davidson1995}
We do not claim the inequalities themselves as new. The questions addressed
here are which partitions of space or of the orbital space make them effective
for molecular dissociation, how they act on the electron-number distribution
of the fragments, and how they combine with the three-index conditions. A
companion paper (Part II)\cite{NakataPartII} treats these inequalities as the spin-summed
two-cell members of a hierarchy of linear programs over integer cell
occupations for the real-space pair density at fixed one-particle density.
For a finite partition the optimum of the program is a lower bound on the
$L^1$ distance of the pair density from the $N$-representable pair densities
with the same one-particle density, and along successively finer partitions
whose cells become arbitrarily small it increases to that distance.
Part II also adds the dual inequalities of these programs as cuts for small
polyatomic systems, linear H$_4^+$ and H$_6^+$ chains and symmetrically
stretched H$_2$O in the STO-6G basis, in which a single region per bond is not
sufficient.

Section~\ref{sec:theory} states the integer electron-count inequalities for
real-space regions and orbital partitions and describes how the regions and
the cuts are selected. Section~\ref{sec:compdetails} gives the systems in the two
basis sets, the relaxations, the region and cut settings, and the solvers. Section~\ref{sec:results} reports the
split-valence calculations of BeB$^+$, LiF, and NO$^+$ and then, in the minimal
basis, the energies, the size consistency at the separated-atom limit, the
fragment charges and dipole moments, and the effect of the axial-symmetry
condition.
Section~\ref{sec:discussion} discusses the two modes of failure, which
partitions are effective, the comparison with subsystem constraints, and the
extension to larger molecules, and Sec.~\ref{sec:conclusions} concludes.

Concurrently with the present work, Boyn imposed the same integer-count
inequalities, equivalent to the bound on the variance of the electron number,
on metal-centered orbital subspaces, but not on real-space regions, in
active-space 2-RDM calculations of mixed-valence transition-metal clusters, in
which the two- and three-index conditions spread redox-active electrons over
several metal centers.\cite{Boyn2026}

\section{Theory}\label{sec:theory}

\subsection{Notation}
We use a finite orthonormal basis of $2n$ spin orbitals built from $n$ real
spatial orbitals $\phi_p$. The first-order RDM is
$\gamma^i_j=\langle\Psi|a_i^\dagger a_j|\Psi\rangle$ with
$\operatorname{Tr}\gamma=N$, and the 2-RDM is
\begin{equation}
  \Gamma^{ij}_{kl}=\tfrac12\langle\Psi|a_i^\dagger a_j^\dagger a_l a_k|\Psi\rangle,
  \qquad \sum_{ij}\Gamma^{ij}_{ij}=\frac{N(N-1)}{2}.
\end{equation}
For an ensemble the expectation values are replaced by ensemble averages. All
conditions below are linear in $(\gamma,\Gamma)$ and are therefore linear
constraints on the variables of the semidefinite program (SDP).

\subsection{Integer electron-count inequalities}
Let $\Nop$ be an operator whose spectrum consists of integers. Because
$(m-k)(m-k-1)\ge0$ for all integers $m$ and $k$,
\begin{equation}
  L_k\equiv\langle(\Nop-k)(\Nop-k-1)\rangle = F-2k\mu+k(k+1)\ge0,
  \label{eq:cut}
\end{equation}
with $\mu=\langle\Nop\rangle$ and $F=\langle\Nop(\Nop-1)\rangle$, holds for
every $N$-electron state and every integer $k$. Inequalities of this form, which
follow from the nonnegativity of a polynomial on the integer spectrum of a
number operator, appear repeatedly in Erdahl's review of representability
conditions.\cite{Erdahl1987} For a count of electrons, whose spectrum is
$\{0,\dots,N\}$, only $k=0,\dots,N-1$ give nontrivial conditions; an integer
combination of counts with signed coefficients has a correspondingly shifted
range.
Because $F=\langle\Nop^2\rangle-\mu$ and
$\operatorname{var}(\Nop)=\langle\Nop^2\rangle-\mu^2$,
\begin{equation*}
  \begin{aligned}
  L_k&=\operatorname{var}(\Nop)+\mu^2-(2k+1)\mu+k(k+1)\\
     &=\operatorname{var}(\Nop)+(\mu-k)(\mu-k-1).
  \end{aligned}
\end{equation*}
Only the second term depends on $k$. It is a parabola in $k$ with its minimum at
$k=\mu-\tfrac12$, so over the integers it is smallest for $k=\lfloor\mu\rfloor$,
\label{rev:floor}where $\lfloor x\rfloor$ ($\lceil x\rceil$) denotes the largest (smallest)
integer not greater (not less) than $x$. The family \eqref{eq:cut} is
therefore equivalent to its member with $k=\lfloor\mu\rfloor$. With
$\theta=\mu-\lfloor\mu\rfloor$ the fractional part of the mean,
$(\mu-k)(\mu-k-1)=\theta(\theta-1)=-\theta(1-\theta)$ for this $k$, and the
member reads
\begin{equation}
  \operatorname{var}(\Nop)\ge\theta(1-\theta),
  \label{eq:var}
\end{equation}
so a fractional mean count must be accompanied by a minimum fluctuation, a bound
that goes back to Yamada.\cite{Yamada1961} The integrality assumption concerns
the spectrum of $\Nop$, that is, the possible outcomes of a single count, and
not its mean: $\mu$ may take any value, but a distribution supported on the
integers with a fractional mean has at least the variance in
Eq.~\eqref{eq:var}, attained when only $\lfloor\mu\rfloor$ and
$\lceil\mu\rceil$ have nonzero weight.

\paragraph{Real-space regions.}
For a region $\Omega$ of the three-dimensional space with sharp boundaries,
let $\Nop$ count the electrons whose positions lie in $\Omega$. Its moments are
integrals of the one-electron density $\rho$ and of the pair density $n_2$,
$\mu=\int_\Omega\rho$ and $F=\iint_{\Omega\times\Omega}n_2$. In the orbital
basis, with the region overlap matrix
$S^\Omega_{pr}=\int_\Omega\phi_p(\bm r)\phi_r(\bm r)\,d\bm r$ placed in both spin
blocks,
\begin{equation}
  \mu=\sum_{pr}S^\Omega_{pr}\gamma^p_r,\qquad
  F=2\sum_{pqrs}S^\Omega_{pr}S^\Omega_{qs}\Gamma^{pq}_{rs}.
  \label{eq:region}
\end{equation}
The matrix $S^\Omega$ is not a projector in a finite basis, but
Eq.~\eqref{eq:region} gives the exact moments of the real-space count of a
finite-basis state, so Eq.~\eqref{eq:cut} is a necessary $N$-representability
condition. The real-space count must not be confused with the one-body operator
$\hat N_S=\sum_{pr}S^\Omega_{pr}a_p^\dagger a_r$ of the finite basis, whose
spectrum is not integer; the two have the same mean, and
$\operatorname{var}(\Nop)=\operatorname{var}(\hat N_S)
+\operatorname{Tr}[(S^\Omega-(S^\Omega)^2)\gamma]$. It is the
member of the Ayers--Davidson family\cite{AyersDavidson2006} with
two-body function $\mathbf 1_\Omega(\bm x)\mathbf 1_\Omega(\bm y)$ and one-body
function $-2k\,\mathbf 1_\Omega$. For fuzzy partitions (for example, Becke or
Hirshfeld weights) the weighted count $\sum_\ell w(\bm r_\ell)$ does not have
an integer spectrum, and the argument above does not apply to it directly; we
use sharp regions only. Placing
$S^\Omega$ in one spin block only gives the corresponding conditions for the
\label{rev:localspin}numbers of $\alpha$ and $\beta$ electrons in $\Omega$. The local spin of a
region is defined in the same real-space sense, with
$\hat S^\Omega_\kappa=\sum_\ell s_\kappa(\ell)\,\mathbf 1_\Omega(\bm r_\ell)$
($\kappa=x,y,z$); $\langle(\hat S^\Omega)^2\rangle$ consists of a two-body
part, evaluated with $S^\Omega\otimes S^\Omega$ and $\Gamma$, and the
one-body part $\tfrac34\operatorname{Tr}(S^\Omega\gamma)$ (supplementary
material).

\paragraph{Orbital partitions.}
If $\{\tilde\phi_i\}$ is an orthonormal set of orbitals with number operators
$\hat n_i$ and $z_i$ are integers, $\hat Q=\sum_i z_i\hat n_i$ has an integer
spectrum and Eq.~\eqref{eq:cut} applies with $\Nop$ replaced by $\hat Q$. In
the localized basis,
\begin{equation}
  L_k=k(k+1)+\sum_i\bigl(z_i^2-(2k+1)z_i\bigr)\tilde\gamma^i_i
      +4\sum_{i<j}z_iz_j\tilde\Gamma^{ij}_{ij}\ge0 ,
\end{equation}
where the indices refer to spin orbitals of the localized basis.

\begin{proposition}\label{prop:twocell}
Let space be divided into two sharp regions $A$ and $B$ and let the total
number of electrons be fixed at $N$. A pair $(\mu,F)$ is the pair of moments
$(\langle\hat N_A\rangle,\langle\hat N_A(\hat N_A-1)\rangle)$ of some
probability distribution of $\hat N_A$ on $\{0,1,\dots,N\}$ if and only if
$0\le\mu\le N$, $L_k\ge0$ for $k=0,\dots,N-1$, and $F\le(N-1)\mu$.
\end{proposition}
\begin{proof}
The attainable pairs form the convex hull of the points
$(m,m(m-1))$, $m=0,\dots,N$, which lie on a convex parabola. The hull is
bounded from below by the segments joining consecutive points, whose
supporting lines are $F=2k\mu-k(k+1)$, $k=0,\dots,N-1$, and from above by the
chord $F=(N-1)\mu$ joining the end points.
\end{proof}
Proposition~\ref{prop:twocell} characterizes the moments of a classical
probability distribution of the count; it is not a statement about the
$N$-representability of $(\gamma,\Gamma)$ as a whole. Every quantum state, pure
or mixed, induces such a distribution, with $p_m$ the probability of finding
$m$ electrons in $A$, so the conditions are necessary in both cases. The
convex hull arises from the convexity of the set of probability distributions,
not from mixing quantum states: for the moments of $\hat N_A$, pure states and
ensembles give the same set, because a pure state
$\Psi=\sum_m c_m\Psi_m$ with $\hat N_A\Psi_m=m\Psi_m$ yields $p_m=|c_m|^2$, and
any distribution is obtained with $c_m=\sqrt{p_m}$ in a Hilbert space that
contains states with every count $m$.

Three of the inequalities in Proposition~\ref{prop:twocell} are implied by the
$P$ condition together with the fixed-$N$ trace and contraction conditions. In real space, $L_0=F=\iint_{A\times A}n_2$,
$L_{N-1}=\langle\hat N_B(\hat N_B-1)\rangle=\iint_{B\times B}n_2$, and
$(N-1)\mu-F=\langle\hat N_A\hat N_B\rangle=\iint_{A\times B}n_2$; all three
are nonnegative because the positive semidefiniteness of $\Gamma$ implies
$n_2(\bm x_1,\bm x_2)\ge0$. The conditions added to an SDP with the $P$
condition are therefore $L_k\ge0$ for $k=1,\dots,N-2$. Among them, the one
with $k=\lfloor\mu\rfloor$ is the most restrictive for a given mean, since it is
equivalent to Eq.~\eqref{eq:var}.
For a two-region partition, the family \eqref{eq:cut} is complete as far as
the first two moments of the spin-summed count are concerned; it is necessary
but not sufficient for the $N$-representability of the 2-RDM. Stronger
conditions require more regions or spin-resolved joint counts, which are
treated in Part II, or higher moments, which involve higher-order RDMs.

\paragraph{Interpretation.}\label{rev:interp}
For two regions at fixed $N$, $\hat N_B=N-\hat N_A$, so that
$\operatorname{cov}(\hat N_A,\hat N_B)=\operatorname{cov}(\hat N_A,N-\hat N_A)
=-\operatorname{var}(\hat N_A)$: every fluctuation of the count of $A$ is
compensated by an opposite fluctuation in $B$. The delocalization index of the
quantum theory of atoms in molecules\cite{Bader1991,Fradera1999} can be written
as $\delta(A,B)=-2\operatorname{cov}(\hat N_A,\hat N_B)$,\cite{MartinPendas2018}
so that Eq.~\eqref{eq:var} becomes $\delta(A,B)\ge2\theta(1-\theta)$: two
fragments with a fractional mean population must share at least a
corresponding number of electrons.
The probabilities $p_m$ themselves, the electron number distribution functions of a
region, can be computed from a wave function and have been analyzed for atomic
partitions of the quantum theory of atoms in molecules, including their
spin-resolved form;\cite{Francisco2007,MartinPendas2007,Francisco2014,BarrenaEspes2024}
Eq.~\eqref{eq:cut} is a condition that a 2-RDM must satisfy for such a
distribution to exist.

\subsection{Selection of regions and cuts}
\label{rev:infinite}Every sharp region, every orthonormal orbital set with integer coefficients
$z_i$, and every integer $k$ give a necessary $N$-representability condition.
The conditions therefore form a continuously infinite family; already for a
fixed orbital set the number of integer vectors $z$ grows exponentially with
the number of orbitals. An SDP can contain only finitely many of them, and we
add only those violated by the current solution, in the manner of a
cutting-plane method. Here the cuts are taken from fixed candidate sets that
are constructed from the Hartree--Fock (HF) solution before any SDP is solved;
the generation of violated conditions by a separation problem instead of
enumeration is treated in Part II.
\begin{enumerate}
\item \emph{Region cuts.} For a diatomic molecule on the $z$ axis with atom
$A$ at the origin, $\Omega=\{z<z_{\mathrm{BCP}}\}$, where $z_{\mathrm{BCP}}$ is
the minimum of the HF density on the bond axis between the nuclei (the bond
critical point); in the split-valence calculations further planes are added
(Sec.~\ref{sec:cutsettings}). For each region the candidate set contains Eq.~\eqref{eq:cut}
for the spin-summed count ($k=0,\dots,N-1$) and for the $\alpha$ and $\beta$
counts. For separated atoms the region is replaced by the projector onto the
orbitals of atom $A$, and for the Hubbard model by the projector onto the sites
of one cluster.
\item \emph{Orbital cuts.}\label{rev:orbcand} The HF orbitals are localized by the Pipek--Mezey
procedure\cite{PipekMezey1989} (meta-L\"owdin populations\cite{SunChan2014})
separately within the doubly occupied, singly occupied, and virtual spaces, so
that the HF determinant is unchanged, and each localized orbital is assigned to
the atom with the largest population. Supports $S$ are sets of localized
spatial orbitals of the following kinds: the orbitals assigned to one atom; the
3, 4, 6, 8, or 10 orbitals with the largest populations on one atom, taken
from all orbitals, from the occupied or the virtual orbitals only, or as a
mixture of both; and sets of the same sizes grown across the bond from pairs of
orbitals of the two atoms. In the STO-6G basis each atom has five and each
diatomic molecule ten spatial orbitals, so the size limit does not exclude any
part of the orbital space; what restricts the candidates is the choice of
subsets by these rules. Each
support gives integer vectors $z$ on its spin orbitals of four kinds: a local
charge ($z_i=1$), a local spin ($z_i=1$ for $\alpha$, $-1$ for $\beta$), a
mixed occupied--virtual combination ($z_i=1$ on occupied and $-1$ on virtual
orbitals), and a charge transfer between the parts of $S$ on the two atoms
($z_i=\pm1$). For each $z$, let $q_{\min}$ and $q_{\max}$ be the smallest and
largest values of $\hat Q$ over determinants with the given numbers of
$\alpha$ and $\beta$ electrons. The candidate set contains
Eq.~\eqref{eq:cut} for every $k$ with $q_{\min}\le k\le q_{\max}-1$; this
range includes the two values adjacent to the HF value of $\hat Q$, which
are the relevant ones as long as the mean stays near its HF value.
Duplicates and sign mirrors [$(z,k)$ and $(-z,-k-1)$ give the same
inequality] are removed. The candidate set contains between \numCandMin{} and
\numCandMax{} candidates per point; the number depends on the numbers of
distinct supports and vectors $z$ and on the ranges of $\hat Q$ (composition
in the supplementary material).
\end{enumerate}
A baseline SDP without cuts is solved first. All candidates violated by the
baseline solution are added as rows of a nonnegative diagonal (linear
programming) block, and the SDP is solved again. The complete candidate set is
then evaluated at the new solution, newly violated candidates are added to
those already present, and the cycle is repeated until no candidate is violated
by more than the tolerance, so that each final solution satisfies the whole
family of cuts of its fixed partition. The tolerance is $10^{-6}$ for the
region cuts and for the orbital cuts of the $P$, $Q$, $G$ relaxation, and
$10^{-4}$ for the orbital cuts of the $P$, $Q$, $G$, $T_1$, $T_2$ relaxation,
for which smaller residual violations are at the level of the solver
precision. The largest remaining violation at the final solutions is
\numFinalRegViolMaxPqg{} ($P$, $Q$, $G$) and \numFinalRegViolMaxTt{} ($P$, $Q$,
$G$, $T_1$, $T_2$) for the region candidates, and \numFinalOrbViolMaxPqg{} and
\numFinalOrbViolMaxTt{}, respectively, for the orbital candidates. Every candidate is evaluated at the HF and full configuration
interaction (FCI) 2-RDMs as a check of validity.

\section{Computational details}\label{sec:compdetails}
\subsection{Systems and reference energies}\label{sec:systems}
\label{rev:svbasis}Two sets of calculations are reported. In the Dunning--Hay
split-valence (SV) basis, with 18 spatial orbitals for each molecule, we
compute the singlet BeB$^+$ (\numdhbebNR{} bond lengths from \numdhbebRmin{} to
\numdhbebRmax~\AA), for which subsystem constraints have been reported in the
same basis,\cite{Verstichel2010} the singlet NO$^+$ (\numdhnoNR{} bond lengths
from \numdhnoRmin{} to \numdhnoRmax~\AA), and the singlet LiF (\numdhlifNR{}
bond lengths from \numdhlifRmin{} to \numdhlifRmax~\AA), whose two lowest
singlet states have an ionic--covalent crossing, each with its
separated-atom limit. In the STO-6G basis, a wider set of systems is computed:
the doublet cations CO$^+$, BeO$^+$, LiF$^+$, and N$_2^+$, the
singlet cation BeB$^+$, the singlet molecules N$_2$, CO, NO$^+$, and CN$^-$
of the N$_2$ isoelectronic series, and the neutral LiF, along a set of bond
lengths and in the separated-atom limit, together
with two four-site Hubbard rings with periodic boundary conditions, coupled
site by site, at half filling and $U/t=4$. The intercluster hopping
$t^\prime$ couples equivalent sites, and $t^\prime/t$ plays the role of the
bond length.

The separated-atom limit is built from separately computed atomic restricted
open-shell HF orbitals with all inter-atomic integrals set to zero.
The HF calculations are restricted (open-shell for the doublets) and use the
\label{rev:hfocc}PySCF program.\cite{Sun2018,Sun2020} The occupation pattern of the HF
orbitals (the electron configuration) of each molecule is fixed along the
bond-length scan to that of the shortest bond length, using a point-group subgroup that fixes the orientation of the $\pi$
orbitals. For NO$^+$, CN$^-$, and N$_2$, whose closed-shell solutions form
crossing branches, the HF solution is followed by continuation from the
near-equilibrium geometry with internal-stability checks. In the SV basis the
HF solution is chosen by the same procedure, and the Hamiltonian is written in
its orbitals. Because the SDP is
formulated in the full orbital space, its solution without cuts does not depend
on the choice of HF orbitals; only the cuts do. For N$_2^+$ and CO$^+$ the STO-6G scans
were repeated with the other low-lying HF configuration
($^2\Sigma_g^+$ and $^2\Pi$, respectively); the results are given in the
supplementary material.

Reference energies are FCI energies. In the SV basis they are those of the
lowest singlet state of $A_1$ symmetry in $C_{2v}$. In the STO-6G basis they
are those of the lowest state with the spin of the
system, obtained with a spin-penalized multi-root Davidson method (penalty
shift $0.5\,E_\mathrm{h}$). Starting with 12 roots and doubling up to 96 when
necessary, the lowest level of roots with the target $\langle\hat S^2\rangle$
(to $10^{-6}$) is accepted only when a higher root with the same spin has been
found, so that the level is complete, and when every root of the level has a
residual below $10^{-6}$ with the unpenalized Hamiltonian; roots within
$10^{-7}\,E_\mathrm{h}$ form one level, of which one root is used. These are
practical criteria for the lowest level, not a proof of its completeness. At
finite bond lengths the FCI levels used are nondegenerate, except the
twofold $\Pi$ levels, whose two components differ only by a rotation about the
bond axis; the real-space properties reported here (charges and counts of
the half-space regions, dipole moments, local spins, and
$\langle\hat L_z^2\rangle$) are invariant under that rotation. The counts of
fixed sets of Pipek--Mezey orbitals in Table~\ref{tab:partitions} need not be;
for the only $\Pi$ state among them (CO$^+$ at 4~\AA), they refer to the
component returned by the eigensolver. At the
separated-atom limit, the FCI levels are degenerate among the multiplet
components of the atoms (and, for N$_2^+$, between the two charge
localizations); there, properties of the FCI root are used only where they are
the same for every root of the level, namely the fragment counts and local
spins of NO$^+$, CN$^-$, CO$^+$, and CO.
Where the Davidson iterations did not converge in reasonable time, which
happened at long bond lengths with nearly degenerate low-lying states, the
FCI Hamiltonian of the $M_S$ sector was diagonalized directly. Both
procedures were applied to one test point, where they agree to within
numerical precision.

\subsection{Relaxations}\label{sec:relaxations}
The SDPs contain the $P$, $Q$, and $G$ conditions and, where stated, the
$T_1$ and $T_2$ conditions,\cite{Erdahl1978,Zhao2004} implemented as in
Ref.~\citenum{Zhao2004}, together with the spin-resolved trace and
contraction conditions. In the SV basis they also contain the singlet
conditions ($S=0$), without symmetry blocking and without the
$\langle\hat L_z^2\rangle$ condition introduced below. With $M_S=0$ alone instead of $S=0$, the
$P$, $Q$, $G$ errors of BeB$^+$ do not reproduce the values of
Ref.~\citenum{Verstichel2010}; with $S=0$ they do (Sec.~\ref{sec:discussion}).
In the STO-6G basis they contain $\langle\hat S_-\hat S_+\rangle=0$, which selects the
\label{rev:ms}component $M_S=S$ of the spin multiplet (with $N_\alpha\ge N_\beta$,
$\langle\hat S_-\hat S_+\rangle=\langle\hat S^2\rangle-M_S(M_S+1)$). For the molecules at finite bond
length, the STO-6G $P$, $Q$, $G$ relaxations, with and without cuts, also contain the
equality $\langle\hat L_z^2\rangle=\Lambda^2$, where $\hat L_z$ is the
component of the electronic orbital angular momentum along the bond axis and
$\Lambda$ is that of the FCI ground state; orbital angular momentum constraints
in variational 2-RDM theory were introduced by Li and
DePrince.\cite{LiDePrince2019} Because $\hat L_z$ is a one-electron
operator, $\langle\hat L_z^2\rangle$ is a linear function of $\gamma$ and
$\Gamma$; in a basis of $s$ and $p$ functions centered on the axis, $\hat L_z$
maps the basis into itself, and the function is exact. For $\Lambda=0$ the
equality characterizes $\Sigma$ states, because $\hat L_z^2$ is positive
semidefinite; for $\Lambda>0$ it is a necessary condition only. The
$P$, $Q$, $G$, $T_1$, $T_2$ relaxations and the separated-atom limits, whose
FCI ground states are degenerate across different $\Lambda$, are computed
without this equality.

\label{rev:gram}Both equalities imply further conditions through the $G$ condition. The $G$
matrix is the Gram matrix $\langle\hat E_{ij}^\dagger\hat E_{kl}\rangle$ of the
one-body operators $\hat E_{kl}=a_k^\dagger a_l$, including the spin-flip
operators $a_{p\alpha}^\dagger a_{q\beta}$. With $v$ the coefficient vector of
$\hat S_+=\sum_p a_{p\alpha}^\dagger a_{p\beta}$, $\langle\hat S_-\hat
S_+\rangle=v^\dagger Gv$; since $G\succeq0$, $v^\dagger Gv=0$ implies $Gv=0$,
and hence $\langle\hat O^\dagger\hat S_+\rangle=0$ for every one-body operator
$\hat O$. The conditions $\langle\hat O\hat S_+\rangle=0$ for one-body $\hat O$
of Ref.~\citenum{VanAggelen2012} are thus implied by the conditions imposed
here. In the same way, $\langle\hat L_z^2\rangle=0$ implies
$\langle\hat O\hat L_z\rangle=0$ for every one-body $\hat O$ at the points
with $\Lambda=0$. The corresponding conditions for two-body $\hat O$ involve
the 3-RDM and are not imposed. Because the equalities hold only to the
tolerance stated below, the implied conditions hold only approximately, with
$\|Gv\|\le(\lambda_{\max}(G)\,v^\dagger Gv)^{1/2}$. For the
\numRdNsol{} solutions of the additional STO-6G calculations described below
(unit coefficients in $v$ and in the vector $u$ of $\hat L_z$),
$v^\dagger Gv\le\numRdSmSpMax$, $\|Gv\|\le\numRdGvMax$,
$\langle\hat L_z^2\rangle\le\numRdLztwoMax$, and $\|Gu\|\le\numRdGuMax$,
with $\lambda_{\max}(G)$ between \numRdLambdaMin{} and \numRdLambdaMax.

\subsection{Regions and cuts}\label{sec:cutsettings}
In the STO-6G basis the region cuts use the half space bounded at the bond
critical point (Sec.~\ref{sec:theory}). In the SV basis they are imposed on
eight half spaces bounded by planes perpendicular to the bond, at the bond critical
point of the HF density and at the fractions $0.2,0.3,\ldots,0.8$ of the bond
length, and, at the separated-atom limit, on the projector onto the orbitals
of one atom. Orbital cuts are generated in both bases as in
Sec.~\ref{sec:theory} from the Pipek--Mezey orbitals of the HF solution, with
the same localization and candidate settings. In both bases violated cuts are
added and the SDP is solved again until no cut of the candidate set is violated.

Region overlap matrices are evaluated on PySCF integration grids
of level 6; region cuts are regarded as violated when $L_k<-10^{-6}$. Because
the half space has a sharp boundary, its overlap matrix does not converge
systematically with the grid; between grid levels the matrix elements vary by
up to \numGridDS{} and the fragment charges by up to \numGridDq~$e$
(Table~S2, STO-6G), which bounds the significance of reported charges. The validity of
the cuts does not depend on this accuracy: for any numerical matrix with
$0\preceq S\preceq I$, the moments of Eq.~\eqref{eq:region} are averages of
the moments of integer orbital counts (include each spin orbital built from an
eigenvector of $S$ independently, with the eigenvalue as probability, so that
every pair of distinct spin orbitals is included with the product of the two
probabilities), so that
Eq.~\eqref{eq:cut} remains a necessary condition. The computed matrices
satisfy this assumption only up to rounding: their eigenvalues leave $[0,1]$ by
at most \numEigExcess. Replacing a matrix by its eigenvalue-clipped version,
for which the argument holds exactly, changes $\mu$ by at most $Nb$ and $F$ by
at most $2b(1+b)N(N-1)$, where $b$ is the excess, so that
$|L_k(S)-L_k(S_\mathrm{c})|\le\delta_k=2b(1+b)N(N-1)+2|k|Nb$, with
$\delta_k$ below \numEigLBound{} for all points. The cuts were built from the
unclipped matrices and imposed as $L_k(S)\ge0$. They are therefore exact
necessary conditions only up to $\delta_k$: every $N$-representable state
satisfies $L_k(S)\ge-\delta_k$, so the imposed condition may exclude such
states by at most this amount. The bound concerns $L_k$, not the energy; the
energy changes caused by the grid are examined directly below.
When the region cuts are rebuilt from the matrices of other grid levels and the
SDPs are solved again, the energies change by less than
\numGridEnergy~m$E_\mathrm{h}$, and by less than \numGridEnergyAbs~m$E_\mathrm{h}$ for
NO$^+$ and CN$^-$ at 6~\AA{} (Table~S2).

\subsection{Solution of the SDPs}\label{sec:solvers}
Equalities are imposed as pairs of inequalities with a tolerance of $10^{-7}$.
All SDPs are written in the dual formulation of the RDM
method:\cite{Zhao2004,Nakata2008} the independent RDM elements are the free
variables, the $N$-representability conditions form a linear matrix inequality
in them, and the linear objective, shifted by the nuclear repulsion, is the
energy. SDPA calls this inequality-form problem the primal problem, and we
follow its convention: we report the energy $E_\mathrm{P}$ of the final
iterate and also record the objective $E_\mathrm{D}$ of the associated
equality-form (dual) problem. For exactly
feasible primal and dual iterates, the optimum $E^\star$ of the relaxation
satisfies $E_\mathrm{D}\le E^\star\le E_\mathrm{P}$. The STO-6G SDPs are solved with SDPA
7.4.3.\cite{Yamashita2003} All of these solutions were
accepted with a relative gap below $10^{-5}$; the absolute gap
$E_\mathrm{P}-E_\mathrm{D}$ has a median of \numGapMedian~m$E_\mathrm{h}$ and a
maximum of \numGapMax~m$E_\mathrm{h}$ over all reported steps, and the gap of
every point is listed in Table~S1; the feasibility errors, equality residuals,
and smallest eigenvalues of representative solutions are given in Table~S4.
Differences smaller than the gap are not
significant. The SV SDPs are solved with an in-house interior-point code, which also returns a
verified lower bound of each relaxation; the bounds lie within the margins stated in
Table~\ref{tab:dhsv}.

\paragraph{Additional STO-6G calculations.}\label{rev:addsto6g}
Two further sets of STO-6G calculations examine the orbital cuts with all
subsets of the localized orbitals as supports (Sec.~\ref{sec:discussion}) and
the position of the plane that bounds the region (Sec.~\ref{sec:discussion}).
They impose the $P$, $Q$, $G$ conditions with
$\langle\hat L_z^2\rangle=\Lambda^2$; the SDPA input files, rebuilt from
the frozen data, are solved with the in-house interior-point code of the
split-valence calculations, here without verified bounds. Violated cuts are added until none is violated, as
above. The energies with the bond-critical-point region and with the
Pipek--Mezey orbital cuts reproduce the SDPA values to within
\numRdZBcpRepro{} and \numRdPmMaxDiff~m$E_\mathrm{h}$, respectively. Of the
\numRdNsol{} solutions, \numRdStatusOpt{} meet the accuracy criterion of the
code; the others stop with the statuses NUMERICAL or dFEAS
(Table~S9), and differences below about 0.1~m$E_\mathrm{h}$ between them are
not interpreted.

\section{Results}\label{sec:results}

\subsection{Split-valence basis}\label{sec:res-sv}
\begin{figure}
\includegraphics[width=\columnwidth]{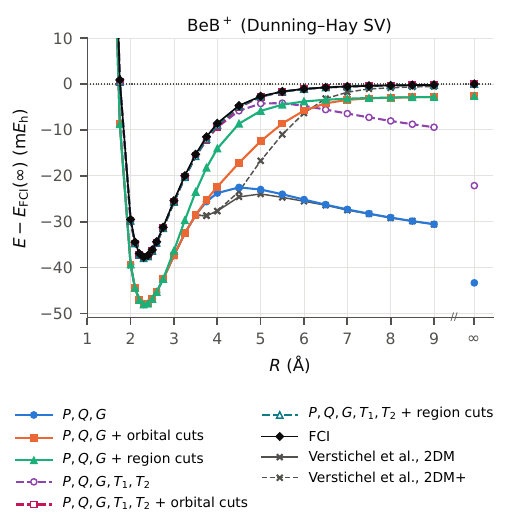}
\caption{Potential energy curves of BeB$^+$ in the Dunning--Hay SV basis, relative to the FCI energy of the separated
atoms: $P$, $Q$, $G$ (solid lines, filled markers) and $P$, $Q$, $G$, $T_1$, $T_2$ (dashed lines, open
markers) without cuts (circles), with orbital cuts (squares), and with region cuts (triangles); FCI in black.
Crosses: $P$, $Q$, $G$ without (2DM) and with (2DM+) subsystem constraints from
Ref.~\citenum{Verstichel2010}, placed at $E_\mathrm{FCI}$ plus the reported errors.
The vertical axis ends 10~m$E_\mathrm{h}$ above $E_\mathrm{FCI}(\infty)$; the errors are listed in
Table~\ref{tab:dhsv}.}
\label{fig:dhsv-beb}
\end{figure}

\begin{figure*}
\begin{minipage}[t]{0.485\textwidth}
\includegraphics[width=\linewidth]{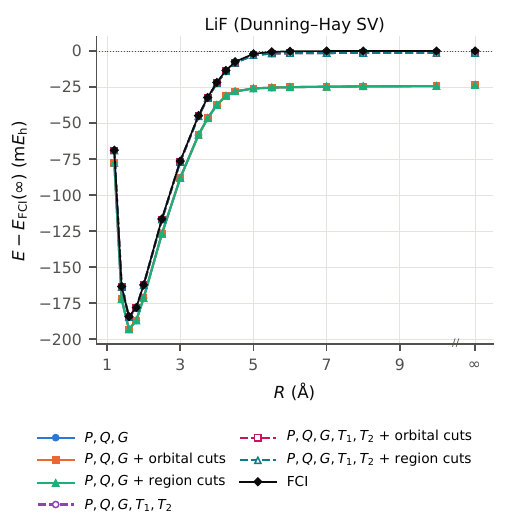}
\caption{Potential energy curves of LiF in the Dunning--Hay SV basis, relative to the FCI energy of the separated
atoms: $P$, $Q$, $G$ (solid lines, filled markers) and $P$, $Q$, $G$, $T_1$, $T_2$ (dashed lines, open
markers) without cuts (circles), with orbital cuts (squares), and with region cuts (triangles); FCI in black.
The vertical axis ends 10~m$E_\mathrm{h}$ above $E_\mathrm{FCI}(\infty)$; the errors are listed in
Table~\ref{tab:dhsv}.}
\label{fig:dhsv-lif}
\end{minipage}\hfill
\begin{minipage}[t]{0.485\textwidth}
\includegraphics[width=\linewidth]{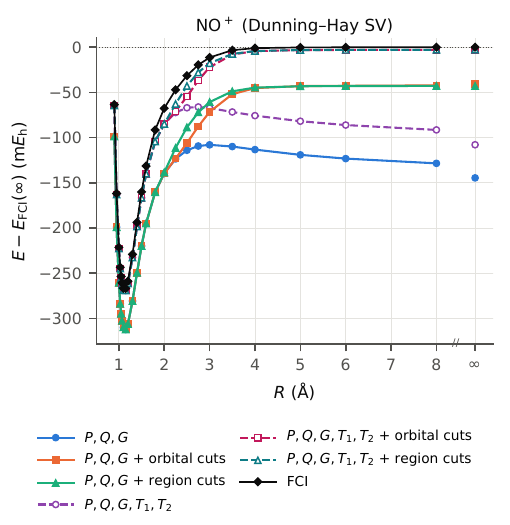}
\caption{Potential energy curves of NO$^+$ in the Dunning--Hay SV basis, relative to the FCI energy of the separated
atoms: $P$, $Q$, $G$ (solid lines, filled markers) and $P$, $Q$, $G$, $T_1$, $T_2$ (dashed lines, open
markers) without cuts (circles), with orbital cuts (squares), and with region cuts (triangles); FCI in black.
The vertical axis ends 10~m$E_\mathrm{h}$ above $E_\mathrm{FCI}(\infty)$; the errors are listed in
Table~\ref{tab:dhsv}.}
\label{fig:dhsv-no}
\end{minipage}
\end{figure*}

\paragraph{Energies.}
Figures~\ref{fig:dhsv-beb}--\ref{fig:dhsv-no} show the potential energy curves
of BeB$^+$, LiF, and NO$^+$ in the Dunning--Hay SV basis, and
Table~\ref{tab:dhsv} lists the errors at selected bond lengths (all points:
Table~S7). For BeB$^+$ and NO$^+$, the error of the $P$, $Q$, $G$ relaxation
grows as the bond is stretched and reaches \numdhbebPQGRinf{} and
\numdhnoPQGRinf~m$E_\mathrm{h}$ at the separated-atom limit. The region cuts
become active from \numdhbebCutOnset{} and \numdhnoCutOnset~\AA, and the
orbital cuts raise the energy by more than 0.1~m$E_\mathrm{h}$ from
\numdhbebOrbOnset{} and \numdhnoOrbOnset~\AA. At long bond lengths the two
partitions give nearly the same energies: \numdhbebPQGorbRNineZeroZero{}
(orbital) and \numdhbebPQGcutRNineZeroZero~m$E_\mathrm{h}$ (region) for BeB$^+$
at 9~\AA, and \numdhnoPQGorbRSixZeroZero{} and
\numdhnoPQGcutRSixZeroZero~m$E_\mathrm{h}$ for NO$^+$ at 6~\AA. At intermediate
bond lengths the region cuts are more effective: \numdhbebPQGorbRFiveZeroZero{}
and \numdhbebPQGcutRFiveZeroZero~m$E_\mathrm{h}$ for BeB$^+$ at 5~\AA, and
\numdhnoPQGorbRThreeZeroZero{} and \numdhnoPQGcutRThreeZeroZero~m$E_\mathrm{h}$
for NO$^+$ at 3~\AA. The errors of NO$^+$ that remain at long bond lengths with
either partition are largely removed by the $T_1$ and $T_2$ conditions; with them, the region cuts keep the errors within
\numdhbebTtRegMax~m$E_\mathrm{h}$ (BeB$^+$) and \numdhnoTtRegMax~m$E_\mathrm{h}$
(NO$^+$) at every bond length. The comparison with the subsystem constraints of
Ref.~\citenum{Verstichel2010} for BeB$^+$ is given in
Sec.~\ref{sec:discussion}.

\paragraph{Ionic--covalent crossing in LiF.}\label{rev:lif}
The two lowest singlet
states of LiF have an avoided crossing near \numdhlifRCross~\AA{} (smallest
computed gap \numdhlifGapMin~m$E_\mathrm{h}$; Li$^+$ + F$^-$ lies
\numdhlifIonicAbove~m$E_\mathrm{h}$ above Li + F at $R=\infty$). The region
cuts are violated from \numdhlifCutOnset~\AA{} on but change the energy by no
more than \numdhlifPqgCutMax~m$E_\mathrm{h}$ with $P$, $Q$, $G$ and
\numdhlifTtCutMax~m$E_\mathrm{h}$ with $P$, $Q$, $G$, $T_1$, $T_2$
(Table~S10), and the orbital cuts by no more than
\numdhlifPqgOrbMax~m$E_\mathrm{h}$ with $P$, $Q$, $G$. The $P$, $Q$, $G$ relaxation without cuts already follows the
transfer of charge: the electron count of the Li half space increases
smoothly from \numdhlifNLiEq{} at 1.6~\AA{} to \numdhlifNLiLong{} at
\numdhlifRmax~\AA, so that no fractional charge is left for the cuts to
correct. Its error, which grows to \numdhlifPqgErrMax~m$E_\mathrm{h}$ in magnitude
beyond the crossing, is reduced to at most \numdhlifTtErrMax~m$E_\mathrm{h}$ by
the $T_1$ and $T_2$ conditions, not by count conditions.

\begin{table*}
\caption{Energy errors $E-E_\mathrm{FCI}$ (m$E_\mathrm{h}$) of BeB$^+$, LiF, and NO$^+$ in the Dunning--Hay SV basis at selected bond lengths (all points: supplementary material), without cuts, with the orbital cuts (orb.) of Sec.~II\,C, and with region cuts (reg.) on the planes bounded at the bond critical point and at fractions $f=0.2$--$0.8$ of the bond length (at $R=\infty$: the atom-orbital projector). Equality conditions: traces, contractions, and the singlet ($S=0$) conditions; no $\langle\hat L_z^2\rangle$ condition. Ref.~\citenum{Verstichel2010}: $P,Q,G$ for BeB$^+$ without (2DM) and with (2DM+) subsystem constraints, as reported there. Every point has a verified lower bound; it lies within 0.53~m$E_\mathrm{h}$ of $E$ (within 0.186~m$E_\mathrm{h}$ for the $P$, $Q$, $G$, $T_1$, $T_2$ solutions with solver status pdOPT). $^a$Solver status other than pdOPT.}
\label{tab:dhsv}
\begin{ruledtabular}
\begin{tabular}{ll*{8}{D{.}{.}{4.3}}}
 & & & & & & & & \multicolumn{2}{c}{Ref.~\citenum{Verstichel2010}} \\
System & $R$ (\AA) & \multicolumn{1}{c}{PQG} & \multicolumn{1}{c}{+orb.} & \multicolumn{1}{c}{+reg.} & \multicolumn{1}{c}{PQGT1T2} & \multicolumn{1}{c}{+orb.} & \multicolumn{1}{c}{+reg.} & \multicolumn{1}{c}{2DM} & \multicolumn{1}{c}{2DM+} \\
\colrule
BeB$^+$ & 1.25 & -16.37 & -16.37 & -16.37 & -1.01 & -1.01 & -1.01 & -16.55 & -16.55 \\
 & 2.30 & -10.32 & -10.32 & -10.32 & -0.260 & -0.260 & -0.260 & -10.30 & -10.30 \\
 & 3.00 & -11.94 & -11.94 & -10.88 & -0.272\rlap{$^a$} & -0.272 & -0.272\rlap{$^a$} & -12.00 & -12.00 \\
 & 4.00 & -15.20 & -13.81 & -5.52 & -0.853 & -0.751 & -0.381 & -19.12 & -19.12 \\
 & 4.50 & -17.83 & -12.48 & -3.94 & -1.13 & -0.395 & -0.193 & -19.90 & -18.73 \\
 & 5.00 & -20.32 & -9.80 & -3.19 & -1.58 & -0.116 & -0.077 & -21.28 & -14.07 \\
 & 6.00 & -24.11 & -4.73 & -2.74 & -3.69 & -0.024 & -0.016 & -24.48 & -5.27 \\
 & 7.00 & -26.79 & -2.93 & -2.66 & -5.91 & -0.010 & -0.008 & -26.95 & -1.26 \\
 & 9.00 & -30.37 & -2.64 & -2.64 & -9.23\rlap{$^a$} & -0.007 & -0.007 & -30.39 & -0.370 \\
 & $\infty$ & -43.33 & -2.54 & -2.64 & -22.15 & -0.006 & -0.009 & \multicolumn{1}{c}{--} & \multicolumn{1}{c}{--} \\
\colrule
LiF & 1.60 & -8.86 & -8.86 & -8.86 & -0.283 & -0.283 & -0.283 & \multicolumn{1}{c}{--} & \multicolumn{1}{c}{--} \\
 & 2.50 & -10.26 & -10.26 & -10.26 & -0.328 & -0.328 & -0.328 & \multicolumn{1}{c}{--} & \multicolumn{1}{c}{--} \\
 & 3.00 & -11.74 & -11.72 & -11.70 & -0.358 & -0.358 & -0.358 & \multicolumn{1}{c}{--} & \multicolumn{1}{c}{--} \\
 & 4.00 & -15.67 & -15.65 & -15.60 & -0.509 & -0.509 & -0.508 & \multicolumn{1}{c}{--} & \multicolumn{1}{c}{--} \\
 & 4.50 & -20.58 & -20.56 & -20.54 & -0.742 & -0.742 & -0.730 & \multicolumn{1}{c}{--} & \multicolumn{1}{c}{--} \\
 & 5.00 & -24.08 & -24.07 & -24.06 & -0.923 & -0.922\rlap{$^a$} & -0.918 & \multicolumn{1}{c}{--} & \multicolumn{1}{c}{--} \\
 & 6.00 & -24.91 & -24.90 & -24.89 & -1.27 & -1.27 & -1.27 & \multicolumn{1}{c}{--} & \multicolumn{1}{c}{--} \\
 & 8.00 & -24.55 & -24.55 & -24.53 & -1.31 & -1.31 & -1.30 & \multicolumn{1}{c}{--} & \multicolumn{1}{c}{--} \\
 & 10.00 & -24.34 & -24.34 & -24.31 & -1.31 & -1.28 & -1.29 & \multicolumn{1}{c}{--} & \multicolumn{1}{c}{--} \\
 & $\infty$ & -23.60 & -23.54 & -23.54 & -1.31 & -1.31 & -1.30 & \multicolumn{1}{c}{--} & \multicolumn{1}{c}{--} \\
\colrule
NO$^+$ & 1.10 & -43.23 & -43.23 & -43.23 & -1.90\rlap{$^a$} & -1.90 & -1.90\rlap{$^a$} & \multicolumn{1}{c}{--} & \multicolumn{1}{c}{--} \\
 & 1.50 & -59.73 & -59.73 & -59.73 & -6.41 & -6.41 & -6.41 & \multicolumn{1}{c}{--} & \multicolumn{1}{c}{--} \\
 & 2.00 & -72.13 & -72.13 & -70.96 & -17.46\rlap{$^a$} & -17.46\rlap{$^a$} & -17.33\rlap{$^a$} & \multicolumn{1}{c}{--} & \multicolumn{1}{c}{--} \\
 & 2.25 & -76.44 & -76.05 & -64.27 & -25.11 & -23.84\rlap{$^a$} & -15.61 & \multicolumn{1}{c}{--} & \multicolumn{1}{c}{--} \\
 & 2.50 & -82.53 & -74.62 & -57.24 & -35.51\rlap{$^a$} & -23.19\rlap{$^a$} & -11.36 & \multicolumn{1}{c}{--} & \multicolumn{1}{c}{--} \\
 & 3.00 & -96.61 & -60.21 & -49.33 & -55.96 & -10.86 & -6.35 & \multicolumn{1}{c}{--} & \multicolumn{1}{c}{--} \\
 & 4.00 & -112.21 & -44.18 & -43.65 & -74.66\rlap{$^a$} & -3.28\rlap{$^a$} & -3.17\rlap{$^a$} & \multicolumn{1}{c}{--} & \multicolumn{1}{c}{--} \\
 & 6.00 & -123.10 & -42.64 & -42.81 & -86.00\rlap{$^a$} & -2.91\rlap{$^a$} & -2.87\rlap{$^a$} & \multicolumn{1}{c}{--} & \multicolumn{1}{c}{--} \\
 & 8.00 & -128.46 & -42.23 & -42.84 & -91.48 & -2.91 & -2.88 & \multicolumn{1}{c}{--} & \multicolumn{1}{c}{--} \\
 & $\infty$ & -144.61 & -40.52 & -42.94 & -107.91\rlap{$^a$} & -2.87 & -2.92 & \multicolumn{1}{c}{--} & \multicolumn{1}{c}{--} \\
\end{tabular}
\end{ruledtabular}
\end{table*}

\subsection{Minimal basis}\label{sec:res-sto}
\begin{figure*}
\includegraphics[width=\textwidth]{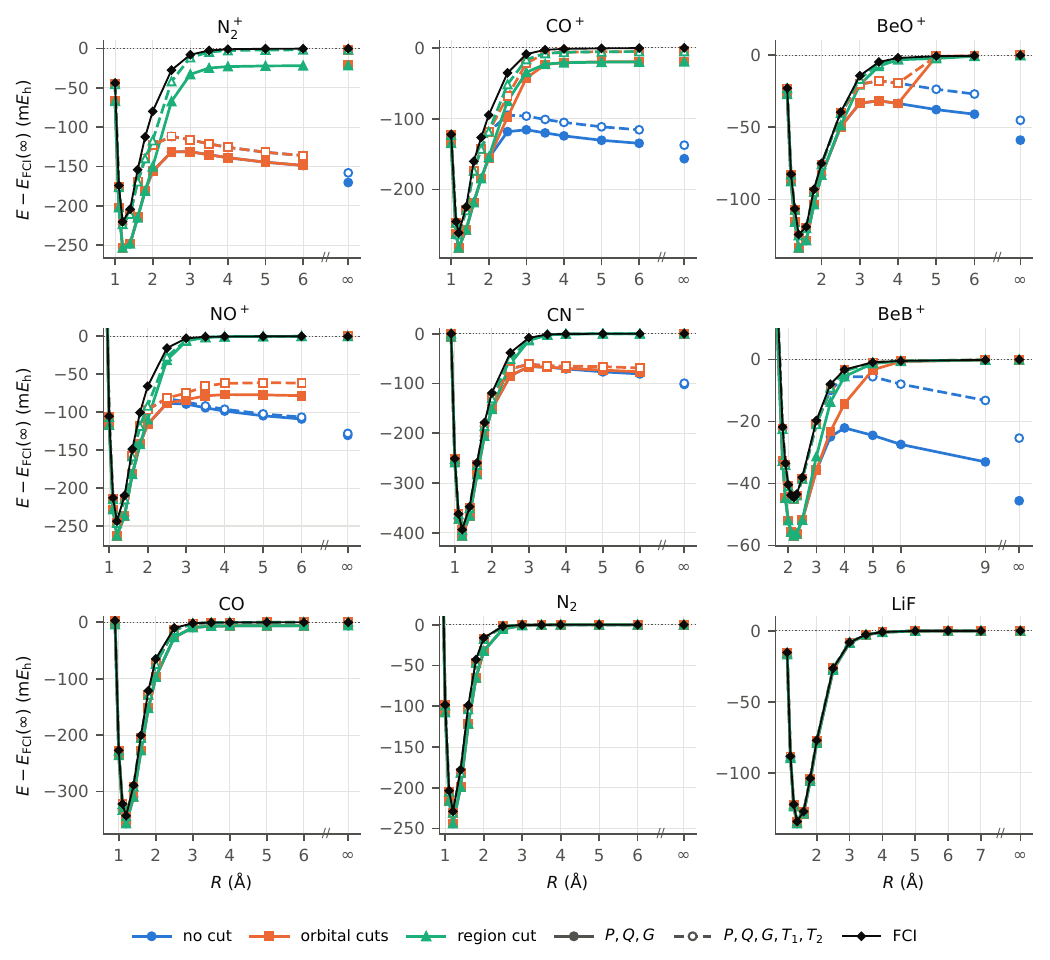}
\caption{Potential energy curves along the bond-length scans (STO-6G),
relative to the FCI energy of the separated atoms, $E_\mathrm{FCI}(\infty)$.
Black diamonds: FCI. Color: no cut, orbital cuts, region cut; solid lines and
filled markers: $P$, $Q$, $G$; dashed lines and open markers: $P$, $Q$, $G$,
$T_1$, $T_2$. The separated-atom limit is shown at the right end of each axis;
the dotted line marks $E_\mathrm{FCI}(\infty)$. The vertical axis ends
10~m$E_\mathrm{h}$ above $E_\mathrm{FCI}(\infty)$, so the repulsive part of a curve
is cut off. Where a cut has no effect, the curve with the cut lies on top of
the curve without cuts and hides it. For BeO$^+$ the orbital cuts take effect only from
5~\AA{} on, because at shorter distances the HF solution from which they are
built has a fractional charge of the Be region (Sec.~\ref{sec:results}).}
\label{fig:pes}
\end{figure*}

\begin{figure*}
\includegraphics[width=\textwidth]{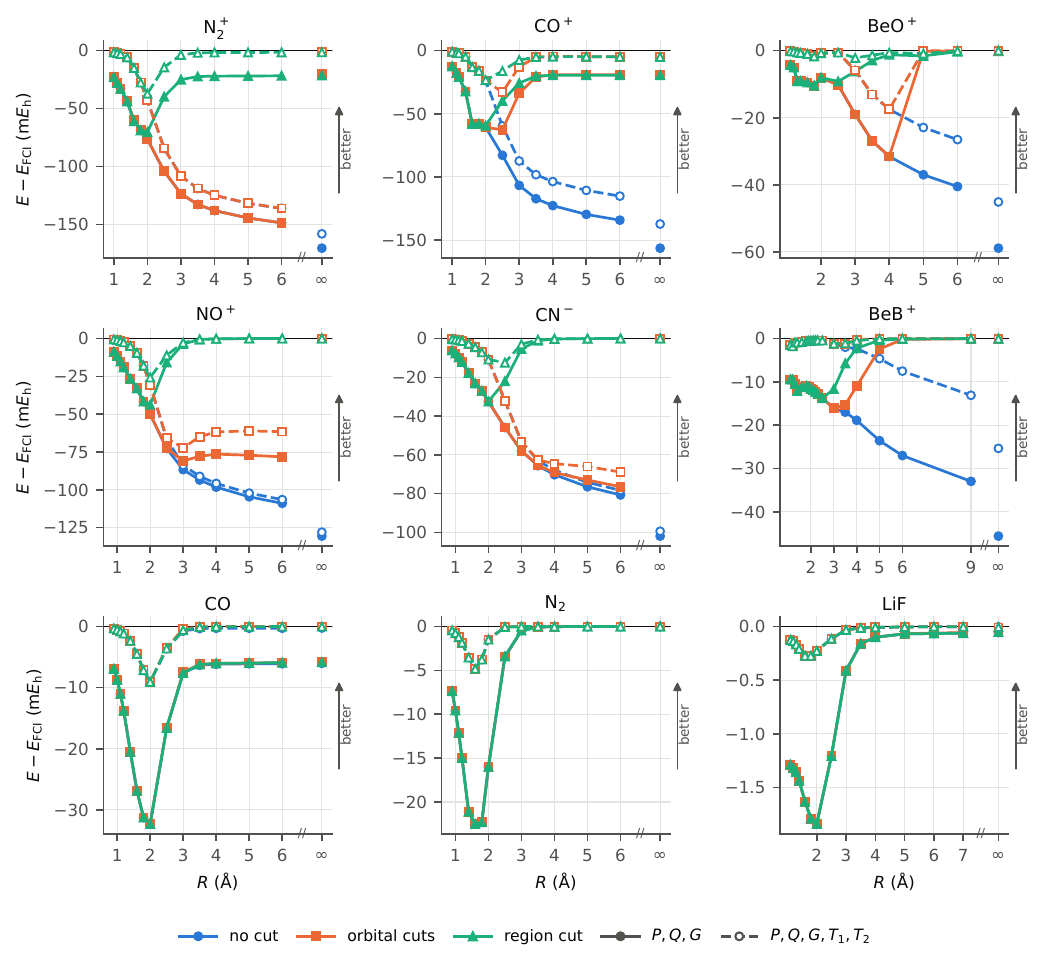}
\caption{Energy errors $E-E_\mathrm{FCI}$ along the bond-length scans (STO-6G).
Color: no cut, orbital cuts, region cut; solid lines and filled markers: $P$,
$Q$, $G$; dashed lines and open markers: $P$, $Q$, $G$, $T_1$, $T_2$. The
separated-atom limit is shown at the right end of each axis. The arrow labeled
``better'' beside each panel points toward the FCI reference. Where a cut has
no effect, as for the neutral molecules, the curve with the cut lies on top of
the curve without cuts and hides it. For BeO$^+$ the orbital cuts take effect only
from 5~\AA{} on, because at shorter distances the HF solution from which they
are built has a fractional charge of the Be region (Sec.~\ref{sec:results}).}
\label{fig:energy}
\end{figure*}

\paragraph{Energies.}
In the STO-6G basis, Figure~\ref{fig:pes} shows the potential energy curves of
all molecules, and
Table~\ref{tab:energies} and Fig.~\ref{fig:energy} compare the energy errors of
the relaxations with and without cuts. For the cations and the anion, the
error of the $P$, $Q$, $G$ relaxation grows with the bond length, and a single
region cut at the bond critical point removes most of it once the bond is
stretched. Orbital cuts built from localized HF orbitals are effective for
some systems, such as CO$^+$ and BeB$^+$ at long bond lengths, but fail for
N$_2^+$ at all finite bond lengths and for NO$^+$ and CN$^-$ at the finite bond
lengths at which the spurious charge transfer appears (Sec.~\ref{sec:discussion}).
For BeO$^+$ the orbital cuts act only from 5~\AA{} on, because the HF solution
used (the lowest over the initial guesses) changes between 4 and 5~\AA{}: its
charge of the Be region is fractional at 4~\AA{} (\numBeOpHfChargeFour{}; FCI
\numBeOpFciChargeFour{}) and integer at 5~\AA{} (\numBeOpHfChargeFive{}), and
only the latter yields violated orbital cuts. The error with orbital cuts
therefore drops from \numBeOpOrbFour{} to \numBeOpOrbFive~m$E_\mathrm{h}$,
whereas the region cut, whose plane also follows the HF density, gives
\numBeOpRegFour{} and \numBeOpRegFive~m$E_\mathrm{h}$ at the two distances.
The $T_1$ and $T_2$ conditions alone reduce these errors only
partly and leave most of the spurious charge transfer in place, while
combining them with the region cut brings the energy close to the FCI value.
For the neutral molecules N$_2$, LiF, and CO the energy errors are smaller and
the cuts change the energies little; CO nevertheless retains a small spurious
charge transfer, discussed below. For the coupled Hubbard rings ($t^\prime>0$)
no cut is violated, and the remaining error is reduced only by the $T_1$ and
$T_2$ conditions; for the uncoupled rings ($t^\prime=0$) a few cuts are
violated, and they change the energy by less than the primal--dual gap.
Table~\ref{tab:energies} shows representative points; all points are listed
in Table~S1.

\paragraph{Size consistency.}
Table~\ref{tab:sizecons} compares, at the separated-atom limit, the energy of
the supersystem with the sum of the energies of the isolated fragments computed
with the same relaxation. The isolated fragments themselves are described almost
exactly: their errors relative to the fragment FCI energies are at most
\numSCFragMax~m$E_\mathrm{h}$. Without cuts, the supersystem energy lies far
below the sum, by \numSCNOpPqg~m$E_\mathrm{h}$ for NO$^+$. With the region or
orbital cuts, the difference $\Delta_\mathrm{SC}$ is at most
\numSCOtherPqgReg~m$E_\mathrm{h}$ ($P$, $Q$, $G$) and \numSCOtherTtReg~m$E_\mathrm{h}$
($P$, $Q$, $G$, $T_1$, $T_2$) in magnitude in the same \numSCRestored{} of the
\numSCPoints{} systems; \numSCNoCutAdditive{} of them (LiF$^+$, N$_2$, LiF) are
nearly additive already without cuts. Numerical intervals built from the
reported primal and dual energies of the supersystem and of the fragments
(Table~S3) contain zero for the region-cut result of \numSCZeroPqg{} systems with the
$P$, $Q$, $G$ conditions and for \numSCZeroTt{} with the $T_1$ and $T_2$
conditions; the corresponding intervals for the remaining systems, including
BeO$^+$, do not contain zero. The intervals would be rigorous bounds only for exactly
feasible primal and dual iterates; the feasibility errors of the solutions
(Table~S4) are small but have not been converted into bounds on the energies. For N$_2^+$ (fragments N, $^4S$, and N$^+$, $^3P$; the dissociated pair
has the energy $E(\mathrm N)+E(\mathrm N^+)$ for either charge localization) and
CO$^+$ (C$^+$, $^2P$, and O, $^3P$), $\Delta_\mathrm{SC}$ remains
\numSCNtwoPPqgReg{} and \numSCCOpPqgReg~m$E_\mathrm{h}$ in magnitude with the
$P$, $Q$, $G$ conditions and \numSCNtwoPTtReg{} and
\numSCCOpTtReg~m$E_\mathrm{h}$ with the $T_1$ and $T_2$ conditions; for CO
(C and O, both $^3P$) it is \numSCCOPqgReg~m$E_\mathrm{h}$ with the $P$, $Q$,
$G$ conditions and \numSCCOTtReg~m$E_\mathrm{h}$ with the $T_1$ and $T_2$
conditions. These residual errors are a lack of additivity of the relaxation,
not errors of the isolated fragments; they are discussed in
Sec.~\ref{sec:discussion}.

\begin{table*}
\caption{Energy errors $E-E_\mathrm{FCI}$ (m$E_\mathrm{h}$; Hubbard model: $10^{-3}t$, coordinate $t^\prime/t$) of the $P,Q,G$ relaxation and of the $P,Q,G,T_1,T_2$ relaxation (PQGT1T2), each without cuts and with the violated orbital (orb.) or region (reg.) cuts added. Equality conditions: traces, contractions, $\langle\hat S_-\hat S_+\rangle=0$; for the $P,Q,G$ columns of the molecules at finite $R$ also $\langle\hat L_z^2\rangle=\Lambda^2$ of the FCI ground state. STO-6G basis.}
\label{tab:energies}
\begin{ruledtabular}
\begin{tabular}{llrrrrrr}
 & & \multicolumn{3}{c}{$P,Q,G$} & \multicolumn{3}{c}{$P,Q,G,T_1,T_2$} \\
System & $R$ (\AA) & PQG & +orb. & +reg. & PQGT1T2 & +orb. & +reg. \\
\colrule
N$_2^+$ & 1.4 & $-43.92$ & $-43.45$ & $-43.92$ & $-5.83$ & $-5.83$ & $-5.83$ \\
 & 3.0 & $-123.44$ & $-123.42$ & $-25.01$ & $-108.28$ & $-108.26$ & $-3.66$ \\
 & 4.0 & $-137.99$ & $-137.99$ & $-22.05$ & $-124.79$ & $-124.78$ & $-1.63$ \\
 & 6.0 & $-148.61$ & $-148.61$ & $-21.73$ & $-136.13$ & $-136.13$ & $-1.50$ \\
 & $\infty$ & $-170.42$ & $-20.26$ & $-21.54$ & $-158.10$ & $-1.43$ & $-1.43$ \\
\addlinespace
CO$^+$ & 1.13 & $-17.64$ & $-17.64$ & $-17.64$ & $-1.60$ & $-1.60$ & $-1.60$ \\
 & 3.0 & $-106.71$ & $-33.76$ & $-25.64$ & $-87.39$ & $-12.68$ & $-7.48$ \\
 & 4.0 & $-122.72$ & $-19.14$ & $-19.52$ & $-103.69$ & $-4.65$ & $-4.74$ \\
 & $\infty$ & $-156.40$ & $-19.13$ & $-19.51$ & $-137.29$ & $-5.02$ & $-5.02$ \\
\addlinespace
BeO$^+$ & 1.4 & $-9.13$ & $-9.13$ & $-9.13$ & $-0.67$ & $-0.67$ & $-0.67$ \\
 & 3.0 & $-19.02$ & $-19.02$ & $-6.25$ & $-5.90$ & $-5.90$ & $-2.17$ \\
 & 4.0 & $-31.57$ & $-31.57$ & $-1.25$ & $-17.50$ & $-17.50$ & $-0.60$ \\
 & $\infty$ & $-58.88$ & $-0.20$ & $-0.20$ & $-45.11$ & $-0.02$ & $-0.02$ \\
\addlinespace
NO$^+$ & 1.1 & $-15.14$ & $-15.14$ & $-15.14$ & $-1.63$ & $-1.63$ & $-1.63$ \\
 & 2.5 & $-72.67$ & $-72.52$ & $-15.82$ & $-66.39$ & $-65.82$ & $-11.06$ \\
 & 4.0 & $-98.20$ & $-76.42$ & $-0.15$ & $-95.76$ & $-61.70$ & $-0.12$ \\
 & 6.0 & $-108.92$ & $-78.19$ & $0.00$ & $-106.45$ & $-61.49$ & $0.00$ \\
 & $\infty$ & $-130.58$ & $0.00$ & $0.00$ & $-127.99$ & $-0.01$ & $0.00$ \\
\addlinespace
CN$^-$ & 1.1 & $-9.73$ & $-9.73$ & $-9.73$ & $-0.82$ & $-0.82$ & $-0.82$ \\
 & 2.5 & $-45.88$ & $-45.88$ & $-22.01$ & $-32.10$ & $-32.09$ & $-12.41$ \\
 & 4.0 & $-70.24$ & $-69.11$ & $-0.22$ & $-67.93$ & $-64.55$ & $-0.15$ \\
 & 6.0 & $-80.67$ & $-76.42$ & $0.00$ & $-78.32$ & $-68.86$ & $0.00$ \\
 & $\infty$ & $-101.89$ & $0.00$ & $0.00$ & $-99.41$ & $0.00$ & $0.00$ \\
\addlinespace
BeB$^+$ & 2.0 & $-11.55$ & $-11.55$ & $-11.55$ & $-0.49$ & $-0.49$ & $-0.49$ \\
 & 4.0 & $-18.85$ & $-11.01$ & $-2.22$ & $-2.33$ & $-0.26$ & $-0.46$ \\
 & 9.0 & $-32.94$ & $-0.03$ & $-0.03$ & $-13.09$ & $0.00$ & $0.00$ \\
 & $\infty$ & $-45.56$ & $-0.03$ & $-0.03$ & $-25.32$ & $0.00$ & $0.00$ \\
\addlinespace
LiF$^+$ & 1.4 & $-1.23$ & $-1.23$ & $-1.23$ & $-0.06$ & $-0.06$ & $-0.06$ \\
 & $\infty$ & $-0.03$ & $0.00$ & $-0.01$ & $-0.01$ & $0.00$ & $0.00$ \\
\addlinespace
CO & 1.1 & $-11.01$ & $-11.01$ & $-11.01$ & $-0.79$ & $-0.79$ & $-0.79$ \\
 & 2.0 & $-32.33$ & $-32.33$ & $-32.33$ & $-9.07$ & $-9.07$ & $-9.07$ \\
 & 6.0 & $-6.09$ & $-5.89$ & $-5.92$ & $-0.27$ & $-0.01$ & $-0.01$ \\
 & $\infty$ & $-6.02$ & $-5.84$ & $-5.86$ & $-0.24$ & $-0.01$ & $-0.01$ \\
\addlinespace
N$_2$ & 1.1 & $-12.12$ & $-12.12$ & $-12.12$ & $-1.22$ & $-1.22$ & $-1.22$ \\
 & 1.6 & $-22.51$ & $-22.51$ & $-22.51$ & $-4.82$ & $-4.82$ & $-4.82$ \\
 & $\infty$ & $0.00$ & $0.00$ & $0.00$ & $0.00$ & $0.00$ & $0.00$ \\
\addlinespace
LiF & 1.6 & $-1.63$ & $-1.63$ & $-1.63$ & $-0.27$ & $-0.27$ & $-0.27$ \\
 & $\infty$ & $-0.06$ & $-0.06$ & $-0.05$ & $0.00$ & $-0.01$ & $-0.01$ \\
\addlinespace
Hubbard 4+4 & 1.5 & $-477.91$ & $-477.91$ & $-477.91$ & $-33.65$ & $-33.65$ & $-33.65$ \\
 & 0.5 & $-448.60$ & $-448.60$ & $-448.60$ & $-48.28$ & $-48.28$ & $-48.28$ \\
 & 0.0 & $-111.86$ & $-111.85$ & $-111.85$ & $0.00$ & $0.00$ & $0.00$ \\
\end{tabular}
\end{ruledtabular}
\end{table*}

\begin{table*}
\caption{Size consistency at the separated-atom limit: $\Delta_\mathrm{SC}=E_{A\cdots B}-(E_A+E_B)$ (m$E_\mathrm{h}$), with the isolated fragments computed with the same relaxation (no cuts are violated for them), the electron number $N_X$ and spin $S_X$ of each fragment taken from the FCI state of the supersystem (N$_2^+$: N and N$^+$), and $M_S=S_X$. Primal energies are used; intervals that include the dual bounds are given in Table~S3. The last column is the largest error of an isolated-fragment relaxation relative to the fragment FCI energy. FCI is additive to $10^{-12}\,E_\mathrm{h}$.}
\label{tab:sizecons}
\begin{ruledtabular}
\begin{tabular}{lllrrrrrrr}
 & & & \multicolumn{3}{c}{$P,Q,G$} & \multicolumn{3}{c}{$P,Q,G,T_1,T_2$} & \\
System & $A$ & $B$ & none & +orb. & +reg. & none & +orb. & +reg. & fragments \\
\colrule
N$_2^+$ & $N=7$, $S=1.5$ & $N=6$, $S=1$ & $-170.42$ & $-20.26$ & $-21.53$ & $-158.10$ & $-1.42$ & $-1.42$ & $0.007$ \\
NO$^+$ & $N=7$, $S=1.5$ & $N=7$, $S=1.5$ & $-130.57$ & $0.00$ & $0.00$ & $-127.99$ & $-0.01$ & $0.00$ & $0.001$ \\
CN$^-$ & $N=7$, $S=1.5$ & $N=7$, $S=1.5$ & $-101.89$ & $0.00$ & $0.00$ & $-99.41$ & $0.00$ & $0.00$ & $0.001$ \\
CO$^+$ & $N=5$, $S=0.5$ & $N=8$, $S=1$ & $-156.39$ & $-19.12$ & $-19.50$ & $-137.28$ & $-5.01$ & $-5.01$ & $0.013$ \\
BeO$^+$ & $N=3$, $S=0.5$ & $N=8$, $S=1$ & $-58.88$ & $-0.19$ & $-0.20$ & $-45.11$ & $-0.01$ & $-0.01$ & $0.002$ \\
BeB$^+$ & $N=4$, $S=0$ & $N=4$, $S=0$ & $-45.56$ & $-0.02$ & $-0.02$ & $-25.32$ & $0.00$ & $0.00$ & $0.003$ \\
LiF$^+$ & $N=2$, $S=0$ & $N=9$, $S=0.5$ & $-0.02$ & $0.00$ & $0.00$ & $0.00$ & $0.00$ & $0.00$ & $0.004$ \\
CO & $N=6$, $S=1$ & $N=8$, $S=1$ & $-6.01$ & $-5.84$ & $-5.85$ & $-0.23$ & $0.00$ & $-0.01$ & $0.005$ \\
N$_2$ & $N=7$, $S=1.5$ & $N=7$, $S=1.5$ & $0.00$ & $0.00$ & $0.00$ & $0.00$ & $0.00$ & $0.00$ & $0.001$ \\
LiF & $N=3$, $S=0.5$ & $N=9$, $S=0.5$ & $-0.05$ & $-0.05$ & $-0.05$ & $0.00$ & $0.00$ & $0.00$ & $0.004$ \\
\end{tabular}
\end{ruledtabular}
\end{table*}

\begin{figure*}
\includegraphics[width=\textwidth]{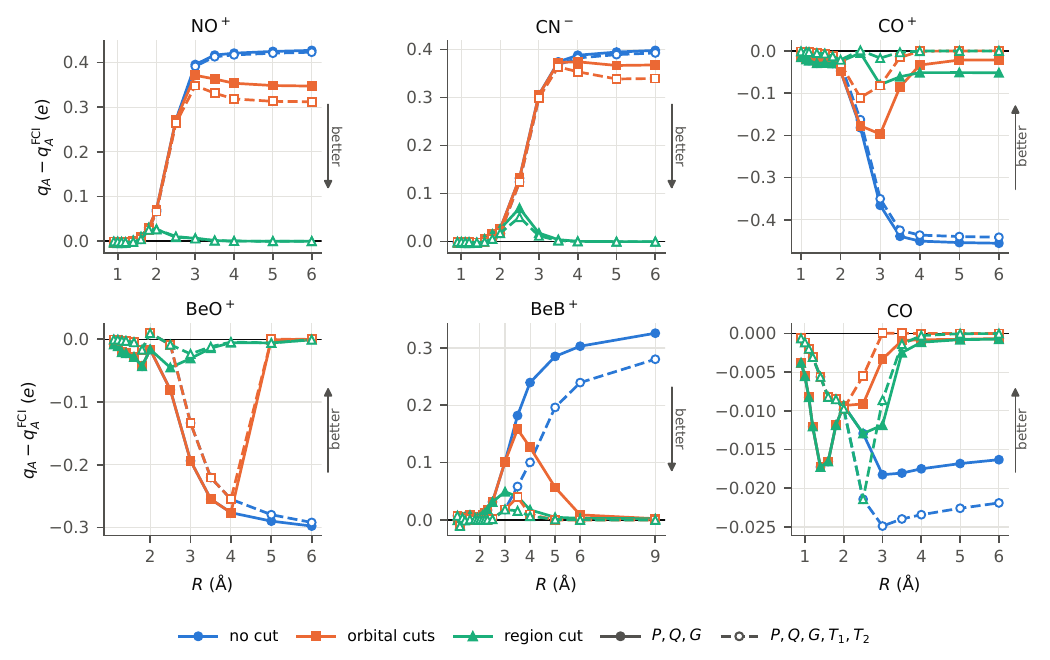}
\caption{Deviation of the charge of the first atom, $q_A-q_A^\mathrm{FCI}$,
in the region $z<z_\mathrm{BCP}$ along the bond-length scans. Encoding as in
Fig.~\ref{fig:energy}; the arrow points toward the FCI value.}
\label{fig:charge}
\end{figure*}

\begin{figure*}
\includegraphics[width=\textwidth]{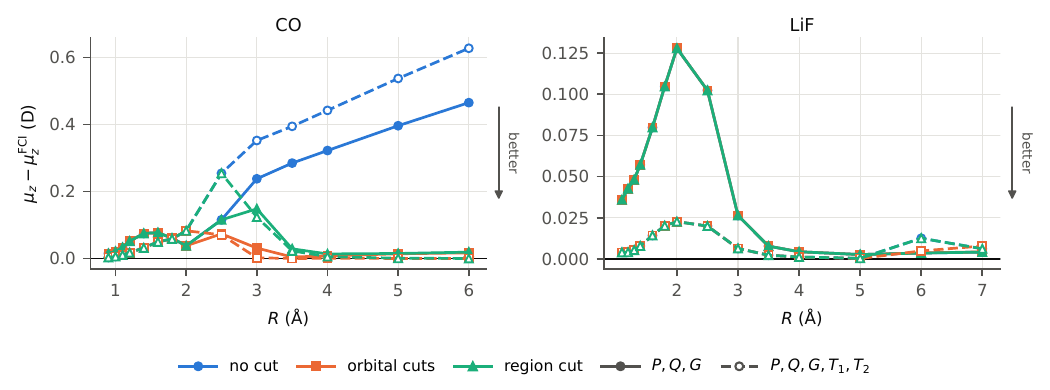}
\caption{Deviation of the dipole moment from the FCI value,
$\mu_z-\mu_z^\mathrm{FCI}$, of neutral CO and LiF along the bond-length scans
($\mu_z>0$: first atom negative). Encoding as in Fig.~\ref{fig:energy}; the
arrow points toward the FCI value. The HF and FCI dipole moments are listed in
Table~S6.}
\label{fig:dipole}
\end{figure*}

\paragraph{Fragment charges and dipole moments.}
Table~S5 and Fig.~\ref{fig:charge} report the charge of the
first atom in the region bounded by the bond-critical-point plane. For the
heteronuclear ions the relaxation without cuts transfers a substantial
fraction of an electron between the separating fragments, and the $T_1$ and
$T_2$ conditions do not remove this transfer; the region cut does. For neutral
CO the spurious transfer is small, but it is independent of the bond length
and therefore produces a dipole moment that grows with the bond length,
whereas the FCI dipole moment vanishes (Table~S6 and
Fig.~\ref{fig:dipole}). The cuts remove most of this dipole moment. The $T_1$
and $T_2$ conditions, which lower the energy error of CO to a fraction of a
milli-hartree at long bond lengths, enlarge the spurious dipole moment. To
separate this effect from that of the axial-symmetry equality, both relaxations
were computed at 6~\AA{} with $\langle\hat L_z^2\rangle=0$: the dipole moment
is \numCOdipPqgLz~D with the $P$, $Q$, $G$ conditions and \numCOdipTtLz~D with
the $T_1$ and $T_2$ conditions; with the region cut it is \numCOdipPqgReg~D
with the $P$, $Q$, $G$ conditions and \numCOdipTtRegLz~D with the $T_1$ and
$T_2$ conditions (FCI: \numCOdipFci~D).

\paragraph{Axial symmetry.}
Without the equality $\langle\hat L_z^2\rangle=\Lambda^2$, the relaxed 2-RDMs
do not in general have the axial symmetry of the FCI ground state. For example,
$\langle\hat L_z^2\rangle$ of the $P$, $Q$, $G$ solution is \numLzNOpFour{} for
NO$^+$ at 4~\AA{} and \numLzCOSix{} for CO at 6~\AA, whereas $\Lambda=0$ in both
cases. Imposing the equality raises the energy by at most
\numLzMaxShift~m$E_\mathrm{h}$ over the \numLzPoints{} points, and by more than
0.5~m$E_\mathrm{h}$ at \numLzNabove{} points, all at short or intermediate bond
lengths (Table~\ref{tab:lambda}); for NO$^+$ at 4~\AA{} and CO at 6~\AA{} the
change is \numLzNOpFourShift{} and \numLzCOSixShift~m$E_\mathrm{h}$. Changes of
the opposite sign, at most \numLzMinShift~m$E_\mathrm{h}$, are within the
precision of the SDP solutions. The fragment charges change by at most
\numLzMaxCharge~$e$. The spurious charge transfer and the large energy errors
at long bond lengths are therefore not consequences of broken axial symmetry.
For the $P$, $Q$, $G$, $T_1$, $T_2$ relaxation, the equality was imposed at
\numLzTtImposed{} representative points only. Compared with the relaxation
without the equality at the same points (\numLzTtPoints{} points), it changes
the energies without cuts and with the region cut by at most
\numLzTtMaxShift~m$E_\mathrm{h}$, but those with the orbital cuts by up to
\numLzTtOrbShift~m$E_\mathrm{h}$, although in both variants the final solutions
satisfy the complete orbital candidate set. The $T_1$, $T_2$ results are
reported without the equality; the region-cut results, on which the
conclusions rest, are insensitive to it. $\Lambda$ is taken from the FCI
ground state, as the target state; in a calculation without a reference, the
symmetry of the target state has to be specified or the candidate symmetries
compared.

\begin{table}
\caption{Effect of the equality $\langle\hat L_z^2\rangle=\Lambda^2$ on the $P,Q,G$ relaxation without cuts: $\Lambda$ of the FCI ground state, $\langle\hat L_z^2\rangle$ of the solution without the equality, and $E-E_\mathrm{FCI}$ (m$E_\mathrm{h}$) without and with it. Listed are the points at which the energy changes by more than 0.5~m$E_\mathrm{h}$.}
\label{tab:lambda}
\begin{ruledtabular}
\begin{tabular}{llrrrr}
System & $R$ (\AA) & $\Lambda$ & $\langle\hat L_z^2\rangle$ & without & with \\
\colrule
N$_2^+$ & 1.0 & 0 & $0.02$ & $-24.59$ & $-23.05$ \\
 & 1.1 & 0 & $0.03$ & $-29.80$ & $-28.01$ \\
 & 1.2 & 0 & $0.05$ & $-35.36$ & $-32.87$ \\
 & 1.4 & 0 & $0.15$ & $-49.35$ & $-43.92$ \\
 & 1.6 & 0 & $0.22$ & $-64.09$ & $-60.96$ \\
 & 1.8 & 0 & $0.23$ & $-70.47$ & $-68.81$ \\
 & 2.0 & 0 & $0.19$ & $-77.15$ & $-76.37$ \\
CO$^+$ & 1.0 & 0 & $0.01$ & $-13.10$ & $-12.56$ \\
 & 1.13 & 0 & $0.02$ & $-18.67$ & $-17.64$ \\
 & 1.2 & 0 & $0.03$ & $-22.36$ & $-20.84$ \\
 & 1.4 & 0 & $0.13$ & $-38.18$ & $-32.34$ \\
 & 1.6 & 1 & $0.39$ & $-62.91$ & $-58.14$ \\
 & 1.8 & 1 & $0.36$ & $-60.93$ & $-57.87$ \\
 & 2.0 & 1 & $0.32$ & $-62.50$ & $-60.74$ \\
BeO$^+$ & 1.2 & 0 & $0.02$ & $-5.65$ & $-5.05$ \\
 & 1.3 & 1 & $0.88$ & $-10.31$ & $-9.13$ \\
 & 1.4 & 1 & $0.94$ & $-9.90$ & $-9.13$ \\
 & 1.8 & 1 & $0.85$ & $-11.20$ & $-10.43$ \\
 & 2.0 & 0 & $0.24$ & $-10.18$ & $-8.06$ \\
 & 2.5 & 0 & $0.14$ & $-11.67$ & $-10.20$ \\
BeB$^+$ & 1.1 & 0 & $1.92$ & $-15.26$ & $-9.42$ \\
 & 1.2 & 2 & $2.24$ & $-16.66$ & $-9.46$ \\
 & 1.3 & 0 & $2.15$ & $-16.65$ & $-10.47$ \\
 & 1.4 & 0 & $1.10$ & $-14.01$ & $-12.07$ \\
CO & 1.8 & 0 & $0.03$ & $-31.82$ & $-31.27$ \\
\end{tabular}
\end{ruledtabular}
\end{table}

\section{Discussion}\label{sec:discussion}

\paragraph{Two modes of failure.}
The calculations show two qualitatively different ways in which the
$P$, $Q$, and $G$ conditions fail in the dissociation of charged diatomic
molecules. Both are failures of the electron-number distribution of a
fragment, and both are removed by the same single-region cut.

In a heteronuclear molecule the charge of each fragment is determined by
the dissociation channel, and the exact state carries an essentially integer
population on each atom at large separation. The relaxed 2-RDM instead
transfers a fraction of an electron between the fragments: the \emph{mean}
population $\mu$ of a region is wrong. This spurious charge transfer is a
fraction of an electron for the cations and anions studied here, persists
to the separated-atom limit, and is not removed by the $T_1$ and $T_2$
conditions. For a neutral heteronuclear molecule it can be small enough to
leave the energy almost unchanged and still produce a dipole moment that
grows with the bond length, whereas the exact dipole moment vanishes.

In a homonuclear cation such as N$_2^+$ the two fragments are equivalent,
and symmetry does not decide which of them carries the positive charge.
The exact state is a superposition of the two charge-localized
configurations, so the mean population of each half is half-integer and is
reproduced by the relaxation, while the count fluctuates between two
integers with variance $\theta(1-\theta)=1/4$. The relaxed 2-RDM keeps the
correct mean but suppresses this fluctuation, a combination that no
probability distribution on integers can produce; it violates
Eq.~\eqref{eq:var}. The same symmetry causes the gerade and ungerade states
to approach each other exponentially with increasing bond length, which
makes the reference calculation itself difficult (Sec.~\ref{sec:compdetails}),
and it makes the symmetry-adapted HF orbitals delocalized over both atoms,
so that orbital cuts built from localized HF orbitals cannot isolate one
fragment.

Table~\ref{tab:moments} shows both modes directly in the moments of the count
of atom $A$. For N$_2^+$ at 4~\AA{} the mean is $6.5$ in all relaxations, but
the variance of the $P$, $Q$, $G$ and of the $P$, $Q$, $G$, $T_1$, $T_2$
solutions is far below the bound $\theta(1-\theta)=1/4$; the region cut
raises it to the bound. For NO$^+$ at 4~\AA{} the mean itself is wrong, and
the variance is again far below the bound for that mean; the region cut
restores the integer mean, whereas the orbital cuts are satisfied by a
moderate increase of the variance with the mean almost unchanged. In both cases
the cut is saturated by the final solution ($L_{\lfloor\mu\rfloor}=0$), and the
violated inequality is the one with $k=\lfloor\mu\rfloor$ for the
bond-critical-point region, that is, Eq.~\eqref{eq:var}.

\begin{table*}
\caption{Moments of the electron count of atom $A$ (bond-critical-point region; at $R=\infty$ the atomic region): mean $\mu$, variance, the bound $\theta(1-\theta)$ of Eq.~(\ref{eq:var}), the slack $L_{\lfloor\mu\rfloor}=\operatorname{var}(\hat N_A)-\theta(1-\theta)$, and the local spins $\langle\hat S_A^2\rangle$ and $\langle\hat S_B^2\rangle$ of atom $A$ and of the complementary region $B$.}
\label{tab:moments}
\begin{ruledtabular}
\begin{tabular}{llrrrrrr}
System & Method & $\mu$ & var & $\theta(1-\theta)$ & $L_{\lfloor\mu\rfloor}$ & $\langle\hat S_A^2\rangle$ & $\langle\hat S_B^2\rangle$ \\
\colrule
N$_2^+$, $R=4.0$~\AA & FCI & $6.500$ & $0.250$ & $0.250$ & $0.000$ & $2.87$ & $2.87$ \\
 & PQG & $6.500$ & $0.001$ & $0.250$ & $-0.249$ & $2.93$ & $2.93$ \\
 & PQG + reg. & $6.500$ & $0.250$ & $0.250$ & $0.000$ & $3.07$ & $3.07$ \\
 & PQGT1T2 & $6.500$ & $0.000$ & $0.250$ & $-0.250$ & $2.80$ & $2.80$ \\
 & PQGT1T2 + reg. & $6.500$ & $0.250$ & $0.250$ & $0.000$ & $2.90$ & $2.90$ \\
\addlinespace
NO$^+$, $R=4.0$~\AA & FCI & $7.000$ & $0.000$ & $0.000$ & $0.000$ & $3.75$ & $3.75$ \\
 & PQG & $6.579$ & $0.000$ & $0.244$ & $-0.243$ & $2.96$ & $2.95$ \\
 & PQG + orb. & $6.647$ & $0.036$ & $0.229$ & $-0.193$ & $3.08$ & $3.08$ \\
 & PQG + reg. & $7.000$ & $0.000$ & $0.000$ & $0.000$ & $3.75$ & $3.75$ \\
 & PQGT1T2 & $6.583$ & $0.000$ & $0.243$ & $-0.243$ & $2.96$ & $2.96$ \\
 & PQGT1T2 + reg. & $7.000$ & $0.000$ & $0.000$ & $0.000$ & $3.75$ & $3.75$ \\
\addlinespace
CO$^+$, $R=\infty$ & FCI & $5.000$ & $0.000$ & $0.000$ & $0.000$ & $0.75$ & $2.00$ \\
 & PQG & $5.462$ & $0.000$ & $0.249$ & $-0.248$ & $1.69$ & $2.75$ \\
 & PQG + orb. & $5.022$ & $0.021$ & $0.021$ & $0.000$ & $1.14$ & $2.04$ \\
 & PQG + reg. & $5.052$ & $0.050$ & $0.050$ & $0.000$ & $1.24$ & $2.09$ \\
 & PQGT1T2 & $5.449$ & $0.000$ & $0.247$ & $-0.247$ & $1.44$ & $2.72$ \\
 & PQGT1T2 + reg. & $5.000$ & $0.000$ & $0.000$ & $0.000$ & $0.92$ & $2.00$ \\
\addlinespace
CO, $R=\infty$ & FCI & $6.000$ & $0.000$ & $0.000$ & $0.000$ & $2.00$ & $2.00$ \\
 & PQG & $6.014$ & $0.001$ & $0.014$ & $-0.013$ & $2.02$ & $2.02$ \\
 & PQG + orb. & $6.000$ & $0.001$ & $0.001$ & $0.000$ & $2.00$ & $2.00$ \\
 & PQG + reg. & $6.001$ & $0.001$ & $0.001$ & $0.000$ & $2.00$ & $2.00$ \\
 & PQGT1T2 & $6.019$ & $0.000$ & $0.018$ & $-0.018$ & $2.03$ & $2.03$ \\
 & PQGT1T2 + reg. & $6.000$ & $0.000$ & $0.000$ & $0.000$ & $2.00$ & $2.00$ \\
\end{tabular}
\end{ruledtabular}
\end{table*}

Once the first two moments of the fragment counts satisfy the conditions of an
integer distribution, the remaining error is of a different kind: it is reduced
by the $T_1$ and $T_2$ conditions and not by further region cuts. The moments
and local spins at the separated-atom limit (Table~\ref{tab:moments}) and the
isolated-fragment calculations (Table~\ref{tab:sizecons}) separate several
contributions. Where the cut removes a spurious charge transfer, it also
restores the local spin; for NO$^+$ at 4~\AA,
$\langle\hat S_A^2\rangle=\numSNOpFourPqg$ without cuts, \numSNOpFourPqgReg{}
with the region cut, and \numSNOpFourFci{} in FCI. For CO$^+$ at the
separated-atom limit with the $P$, $Q$, $G$ conditions, the atomic-region cut
does not make the count of C$^+$ integer: the mean is \numMCOpPqgRegMu{} and
the variance \numMCOpPqgRegVar{}, exactly the integer bound for that mean;
the residual \numSCCOpPqgReg~m$E_\mathrm{h}$ thus coexists with a small
fractional mean population whose variance is at the integer-count bound. With
the $T_1$ and $T_2$ conditions the count becomes integer to the displayed
precision (\numMCOpTtRegMu) and the local spin of O is correct (\numMCOpTtRegSB), but that
of C$^+$ is \numMCOpTtRegSA{} instead of \numMCOpFciSA, while the residual is
\numSCCOpTtReg~m$E_\mathrm{h}$. For CO with the $P$, $Q$, $G$ conditions, in
contrast, the count is nearly integer (\numMCOPqgRegMu) and both local spins
are nearly correct (\numMCOPqgRegSA{} and \numMCOPqgRegSB), and so is the
spin correlation $\langle\hat{\bm S}_A\cdot\hat{\bm S}_B\rangle
=\tfrac12(\langle\hat S^2\rangle-\langle\hat S_A^2\rangle-\langle\hat S_B^2\rangle)$
of the total singlet; yet $\Delta_\mathrm{SC}$ is
\numSCCOPqgReg~m$E_\mathrm{h}$. These observables therefore do not identify
the remaining defect of the relaxed 2-RDM. In N$_2^+$ at 4~\AA{} the region cut restores the variance of the
count but raises the $P$, $Q$, $G$ value of $\langle\hat S_A^2\rangle$ from
\numSNtwoPlusFourPqg{} to \numSNtwoPlusFourPqgReg, compared with
\numSNtwoPlusFourFci{} in FCI; adding the $T_1$ and $T_2$ conditions brings it
to \numSNtwoPlusFourTtReg.

These observations are consistent with known limitations of approximate
$N$-representability conditions for states with spin. That the $M_S$
components of a multiplet are not degenerate under these conditions was
reported for the $^3P$ states of C and O, with the $P$, $Q$, $G$ conditions and
with the $T_1$ and $T_2$ conditions, together with its consequence for size
consistency in dissociation;\cite{Nakata2008} the component with $M_S=S$ is the
most strongly constrained, and correct spin expectation values of the
supersystem do not imply correct spins of the dissociation
products;\cite{VanAggelen2012} size-consistency errors in dissociation also
depend on the spin and orbital-angular-momentum projections of the
fragments.\cite{LiDePrince2019} Our isolated fragments are computed with
$M_S=S$, whereas in the supersystem their spins are coupled to the total spin:
for two $S=1$ fragments coupled to a singlet, the spin state of each fragment
is the equal mixture of its three $M_S$ components, with
$\langle\hat S_{Az}\rangle=0$ but $\langle\hat S_{Az}^2\rangle=2/3$, not the
$M_S=0$ component alone. The
data show that the residual nonadditivity of CO$^+$ is accompanied by an
incorrect local spin of a fragment whose count is correct, but they do not
establish the local spin as its cause, and for CO the residual occurs with
nearly correct local spins. Conditions on the spin structure of the fragments
beyond these expectation values are therefore natural complements to the
electron-count cuts; they
are not studied here.

\paragraph{Which partitions are effective.}
The region cut and the orbital cuts use the same inequality,
Eq.~\eqref{eq:cut}, but differ in the partition, and in the minimal basis the
partition strongly affects whether the cut is effective. Table~\ref{tab:partitions} compares, at
\numPartPoints{} points, the bond-critical-point region and the Pipek--Mezey orbital cuts with
partitions defined by the atomic orbitals of atom $A$ orthogonalized by the
L\"owdin,\cite{Lowdin1950} meta-L\"owdin,\cite{SunChan2014} and
intrinsic-atomic-orbital (IAO)\cite{Knizia2013} procedures; the IAOs are
built from all HF orbitals with the default MINAO reference basis of PySCF, so
that in the minimal basis they span the whole orbital space, and are
orthogonalized symmetrically; meta-L\"owdin orbitals use the ANO pre-orthogonalization
of PySCF. All atomic-orbital partitions give the same
energies as the bond-critical-point region to within
\numPartMaxDiff~m$E_\mathrm{h}$, each with a single violated cut, whereas the
Pipek--Mezey orbital cuts are ineffective for N$_2^+$, NO$^+$, and CN$^-$ in the
STO-6G basis. The
\label{rev:pmall}restriction of the supports is not the cause: with every nonempty subset of the
ten localized orbitals admitted as a support (\numRdPmCandMin--\numRdPmCandMax{}
candidates), the errors are \numRdPmNtwopFourErr{} (N$_2^+$, 4~\AA),
\numRdPmNOpSixErr{} (NO$^+$, 6~\AA), and \numRdPmCNmSixErr~m$E_\mathrm{h}$
(CN$^-$, 6~\AA), compared with \numRdPmNtwopFourPartIErr, \numRdPmNOpSixPartIErr,
and \numRdPmCNmSixPartIErr~m$E_\mathrm{h}$ with the candidate set of
Sec.~\ref{sec:theory}. In the STO-6G basis the ineffectiveness is therefore not
caused by the selection of supports. It is not a general property of the
Pipek--Mezey orbital cuts either: in the split-valence basis the orbital cuts of
NO$^+$ are nearly as effective as the region cut at long bond lengths, although
less effective at intermediate ones (Table~\ref{tab:dhsv}). The relative
performance of the two partitions thus depends on the basis and on the bond
length, and the present calculations do not establish a general rule. A
possible indicator, in the systems studied, is the slack
$L_{\lfloor\mu\rfloor}=\operatorname{var}(\hat N_A)-\theta(1-\theta)$ of the condition
at the FCI state. For the atomic-orbital partitions and the
bond-critical-point region at long bond lengths, this slack is close to zero:
the count of atom $A$ is an integer with negligible variance or, in N$_2^+$,
has a half-integer mean with exactly the minimal variance
$\theta(1-\theta)=1/4$. A relaxed 2-RDM with a
wrong fractional mean can then satisfy the inequality only by acquiring the
variance $\theta(1-\theta)$. A plausible interpretation, not a proof for
relaxed 2-RDMs, is that such a variance requires ionic configurations of high
energy, so that correcting the mean is the cheaper way to satisfy the
condition. The Pipek--Mezey
orbitals are localized separately in the occupied and virtual spaces and keep
small tails on the other atom. For NO$^+$ at 6~\AA, the count of the orbitals
assigned to N has variance \numTNOpPmFciVar{} at the FCI state; the relaxed
2-RDM, with mean \numTNOpPmPqgMu{} and variance \numTNOpPmPqgVar, satisfies the
added cut by increasing the variance to \numTNOpPmOrbVar{} with nearly the same
mean (\numTNOpPmOrbMu), and the L\"owdin count remains violated
($L_k={}$\numTNOpLowdinOrbL). For N$_2^+$ the symmetry-adapted HF orbitals cannot
be localized on one atom within the occupied and virtual spaces separately, and
no orbital set represents one fragment. For CO$^+$ at 4~\AA{} the
slack of the Pipek--Mezey count at the FCI state is close to zero, and the
orbital cuts are as effective as the region cut. A small slack at the exact
state is not sufficient by itself: the condition must also be violated by the
relaxed solution, and how much the energy changes when the violation is removed
depends on the Hamiltonian.

\begin{table*}
\caption{Energy errors $E-E_\mathrm{FCI}$ (m$E_\mathrm{h}$) of the $P,Q,G$ relaxation without cuts and with the violated cuts of different partitions: Pipek--Mezey HF orbitals (orb.), the bond-critical-point region (BCP), and the atomic orbitals of atom $A$ orthogonalized by the L\"owdin, meta-L\"owdin, and intrinsic-atomic-orbital (IAO) procedures. The last two columns give the slack $L_k$, $k=\lfloor\mu\rfloor$, of the integer-count inequality for atom $A$ at the FCI state, for the Pipek--Mezey orbitals assigned to $A$ and for the L\"owdin orbitals of $A$.}
\label{tab:partitions}
\begin{ruledtabular}
\begin{tabular}{llrrrrrrrr}
 & & & \multicolumn{5}{c}{$E-E_\mathrm{FCI}$ with cuts} & \multicolumn{2}{c}{$L_k$ at FCI} \\
System & $R$ (\AA) & none & orb. & BCP & L\"owdin & meta-L\"owdin & IAO & orb. & L\"owdin \\
\colrule
N$_2^+$ & 4.0 & $-137.99$ & $-137.99$ & $-22.05$ & $-21.90$ & $-21.90$ & $-21.95$ & $0.686$ & $0.000$ \\
CO$^+$ & 4.0 & $-122.72$ & $-19.14$ & $-19.52$ & $-19.40$ & $-19.40$ & $-19.39$ & $0.000$ & $0.000$ \\
NO$^+$ & 6.0 & $-108.92$ & $-78.19$ & $0.00$ & $0.00$ & $0.00$ & $0.00$ & $0.046$ & $0.000$ \\
CN$^-$ & 6.0 & $-80.67$ & $-76.42$ & $0.00$ & $0.00$ & $0.00$ & $0.00$ & $0.106$ & $0.000$ \\
\end{tabular}
\end{ruledtabular}
\end{table*}

\paragraph{Comparison with subsystem constraints.}\label{rev:subsys}
Verstichel \emph{et al.}\ studied the singlet BeB$^+$ with the $P$, $Q$, and
$G$ conditions in a Dunning--Hay basis and reported energy errors relative
to FCI of 30.39~m$E_\mathrm{h}$ at a bond length of 9~\AA, reduced to
0.37~m$E_\mathrm{h}$ by subsystem constraints.\cite{Verstichel2010}
Figure~\ref{fig:dhsv-beb} and Table~\ref{tab:dhsv} compare these values with the
region cuts in the Dunning--Hay SV basis. Without cuts, our $P$, $Q$, $G$
errors agree with those of Ref.~\citenum{Verstichel2010} within
\numdhbebRefDevMax~m$E_\mathrm{h}$ at all bond lengths except
\numdhbebRefDevFrom--\numdhbebRefDevTo~\AA, where the reported errors are
larger in magnitude by up to \numdhbebRefDevMid~m$E_\mathrm{h}$; the agreement
elsewhere is consistent with the same basis, and we have not identified the
cause of the difference. With the region cuts, the $P$, $Q$, $G$ error at
9~\AA{} is \numdhbebPQGcutRNineZeroZero~m$E_\mathrm{h}$, compared with
\numdhbebRefPlusNine~m$E_\mathrm{h}$ with subsystem constraints. From 3 to
6~\AA{} the region cuts give the smaller errors (at 4.5~\AA{}
\numdhbebPQGcutRFourFiveZero{} versus \numdhbebRefPlusFourFive~m$E_\mathrm{h}$;
part of this difference is already present without cuts), and from 6.5~\AA{}
on the subsystem constraints do. This is expected from the
form of the conditions: at large separation, a subsystem constraint bounds
the energy of each fragment and thus restricts the fragment state, whereas a
count inequality bounds only the variance of the fragment electron number.
Combined with $T_1$ and $T_2$, the region cuts reduce the error of BeB$^+$ to
at most \numdhbebTtRegMax~m$E_\mathrm{h}$ in magnitude at every bond length,
and to \numdhbebTOneTwocutRNineZeroZero~m$E_\mathrm{h}$ at 9~\AA. At the
separated-atom limit, the Be charge is \numdhbebChargePqg~$e$ without and
\numdhbebChargePqgReg~$e$ with the cut for $P$, $Q$, $G$, and
\numdhbebChargeTt{} and \numdhbebChargeTtReg~$e$ for $P$, $Q$, $G$, $T_1$,
$T_2$; Ref.~\citenum{Verstichel2010} reported Mulliken charges of $+0.38$ and
$0.00$ at 9~\AA{} without and with subsystem constraints.

The two approaches differ in what they
require: a subsystem constraint needs
separate variational calculations of the isolated subsystem for several
electron numbers, whereas the cuts used here are Hamiltonian independent
and are constructed from the HF density alone. For orbital partitions, the
count inequalities can also be viewed as subsystem bounds for the auxiliary
Hamiltonian $\hat N_A(\hat N_A-1)$, which depends only on the electron number
of the subsystem and whose energies $m(m-1)$ are known analytically; the convex
interpolation between adjacent integers gives $F\ge2k\mu-k(k+1)$, which is
Eq.~\eqref{eq:cut}. Their practical advantage is that no variational
calculations of the physical fragment Hamiltonians are required. For subsystem
constraints, the atomic data can be reused for a given basis, whereas
multi-atom subsystems require geometry-dependent calculations.

\paragraph{Choice of the region boundary.}\label{rev:boundary}
The plane through the bond critical point of the HF density is a simple
approximation of the zero-flux surface of the quantum theory of atoms in
molecules, which is curved for heteronuclear molecules. The validity of the
cuts does not depend on this choice: every sharp region, of any shape, gives
necessary conditions. Their effectiveness does, because a condition acts only
if the exact count of the region is nearly integer or nearly saturates
Eq.~\eqref{eq:var}. Table~S9 lists the STO-6G energies with the plane at
nine positions between $0.30R$ and $0.70R$, the first atom being at $z=0$. At
4~\AA{} the position matters little: for $0.35R\le z_0\le0.55R$ the errors of
NO$^+$ and CN$^-$ are at most \numRdZNOpFourMaxErrInner{} and
\numRdZCNmFourMaxErrInner~m$E_\mathrm{h}$ in magnitude. At shorter bond lengths
the bond critical point is not always the best position. For NO$^+$ at 2.5 and
3.0~\AA, the planes at $z_0=\numRdZNOpTwoFiveBestZ R$ and
$\numRdZNOpThreeBestZ R$, closer to N than the bond critical point ($\numRdZNOpTwoFiveBcp R$ and $\numRdZNOpThreeBcp R$), reduces the
error from \numRdZNOpTwoFiveBcpErr{} to \numRdZNOpTwoFiveBestErr~m$E_\mathrm{h}$
and from \numRdZNOpThreeBcpErr{} to \numRdZNOpThreeBestErr~m$E_\mathrm{h}$;
for CN$^-$ at 2.5~\AA{} and for N$_2^+$ at 4~\AA{} the bond critical point is the
best of the positions examined. Imposing all planes at once gives energies within
\numRdZAllGapMax~m$E_\mathrm{h}$ of the best single plane: within the family
of half spaces, the position of one plane matters, not the number of planes.
This agrees with the convexity of the set of $N$-representable 2-RDMs, for
which one valid inequality, $\langle\hat H\rangle\ge E_\mathrm{FCI}$, already
makes any relaxation exact; a restricted family such as Eq.~\eqref{eq:cut}
need not contain such an inequality, and adding members helps only insofar
as they combine into a better one. Since the plane at every position gives a
valid lower bound, the position can be chosen without the FCI reference as the
one that gives the highest energy. A golden-section search over
$0.35R\le z_0\le0.55R$ gives the errors \numRdZNOpTwoFiveOptErr{}
($z_0=\numRdZNOpTwoFiveOptZ R$) and \numRdZNOpThreeOptErr~m$E_\mathrm{h}$
($\numRdZNOpThreeOptZ R$) for NO$^+$ at 2.5 and 3.0~\AA, and
\numRdZCNmTwoFiveOptErr{} ($\numRdZCNmTwoFiveOptZ R$) and
\numRdZCNmThreeOptErr~m$E_\mathrm{h}$ ($\numRdZCNmThreeOptZ R$) for CN$^-$, compared
with \numRdZCNmTwoFiveBcpErr{} and \numRdZCNmThreeBcpErr~m$E_\mathrm{h}$ at the
bond critical point; near the optimum the energy varies by only a few
$10^{-2}$~m$E_\mathrm{h}$, so the optimal positions are determined only to
about $0.02R$. For a boundary that depends on parameters
$p$, the derivative of the optimal energy follows from one solution,
$\mathrm{d}E/\mathrm{d}p=-\eta\,\partial L_k(\Gamma^\star;p)/\partial p$, where
$\Gamma^\star$ is the solution and $\eta\ge0$ the multiplier of the active
cut, whenever the optimum depends differentiably on $p$; the shape of the
region can therefore be optimized by gradient methods, although $E(p)$ need not
be concave and the maximum found need not be global. The
split-valence calculations impose eight planes
simultaneously (Sec.~\ref{sec:cutsettings}). At the four STO-6G points of Table~\ref{tab:partitions}, the
atomic-orbital partitions, which are not bounded by planes, give energies
within \numPartMaxDiff~m$E_\mathrm{h}$ of the plane. Near an avoided ionic--covalent crossing, as in
the alkali halides, the exact count of a region is not close to an integer
for any boundary, and no count condition can be nearly saturated; the cuts
then remain valid but are expected to be weak. The STO-6G calculations of LiF
do not test this case, because the FCI ground state remains covalent at all
bond lengths computed. In the split-valence basis (Sec.~\ref{sec:res-sv}) the
cuts on LiF are indeed weak, but because the relaxation without cuts already
follows the charge transfer through the crossing, so that the error is not
caused by a fractional fragment charge.

\paragraph{Polyatomic molecules.}
The half space bounded at the bond critical point is specific to diatomic
molecules, but the atomic-orbital partitions of Table~\ref{tab:partitions}
are not. For any molecule, the sum of the number operators of the
orthogonalized atomic orbitals of an atom $A$, $\hat N_A=\sum_{\mu\in A}\hat
n_\mu$, has an integer spectrum, so Eq.~\eqref{eq:cut} applies to it and to
sums over several atoms without a numerical integration grid or a search for
critical points. Evaluating all candidate cuts at a given solution is
inexpensive, and only the violated ones are added to the SDP. The present
examples suggest giving priority to regions for which the condition is nearly
saturated by the exact state but violated by the relaxed solution: regions whose exact count is
close to an integer but whose relaxed mean is fractional, such as the fragments
separated by a stretched bond, and regions whose mean is correct but whose
variance is suppressed below the integer bound, as in N$_2^+$. Small polyatomic systems, for which a single region
per bond is not sufficient, are treated in Part II.

\paragraph{Number of cuts for larger molecules.}\label{rev:largemol}
For a molecule with $M$ atoms partitioned into atomic regions, the
single-region family \eqref{eq:cut} contains $O(MN)$ candidates if all
nontrivial $k$ and the three spin channels are included. Only the
inequalities with $k$ close to the mean population of a region can become
active, so restricting $k$ to a few integers around the HF population of each
region leaves $O(M)$ candidates; more simply, for each region only the
inequality with $k=\lfloor\mu\rfloor$ of the current solution needs to be
evaluated, because it is the most restrictive one (Sec.~\ref{sec:theory}). The number of
candidates examined per iteration must be distinguished from the total number
of cuts added over all iterations. Fragments made of several atoms enlarge the
candidate set combinatorially if all subsets are admitted. In molecules,
however, the number of neighbors of an atom does not grow with the size of the
molecule, and electron-number fluctuations between distant fragments are
\label{rev:benzene}usually small. This does not hold for delocalized systems; in benzene, for
example, the delocalization index of carbon atoms in para positions is
larger than that of atoms in meta positions. Delocalization indices computed
at the HF level could serve to select the multi-atom fragments that need
candidates; restricting multi-atom fragments to connected sets of bonded atoms with a
number of atoms bounded independently of $M$ keeps the number of candidates at
$O(M)$, whereas admitting all connected fragments does not (already for a chain,
the contiguous segments number $O(M^2)$). We expect the number of active cuts to be of the order
of the number of fragments that carry a fractional population, that is, of the
number of stretched bonds. This estimate is extrapolated from the diatomic
calculations reported here. In the linear H$_6^+$ chain of Part II, cuts on
single atomic regions leave the error unchanged, and the violations are carried
by fragment regions of several atoms (pairs of neighboring atoms, halves of the
chain, and interior triples), so the fragments that need cuts are not always
single atoms. The estimate is also not expected to hold for systems with
strongly delocalized electrons, in which the counts of distant regions remain
correlated.
Joint distributions of the counts of several regions involve a number of
integer configurations that grows rapidly with the number of regions; such
conditions are generated from the current solution by separation rather than
enumerated in advance (Part II).

\paragraph{Cost per cut.}
Each cut adds one row to a nonnegative diagonal block and does not change the
semidefinite blocks. In the delocalized molecular-orbital basis used here, the
row of a region cut has nonzero coefficients for almost all 2-RDM variables
($O(n^4)$ for $n$ spatial orbitals). In a localized basis, the coefficients
decay with the distance of the orbitals from the region; Gaussian orbitals have
tails, so the coefficients are not strictly zero, but if those below a
threshold are neglected, with a controlled error, the number of retained
coefficients is expected to become independent of the size of the molecule
for regions made of a bounded number of atoms, with a fixed basis per atom and
orbitals that remain localized; this does not apply to large regions, inside
which the diagonal elements of $S^\Omega$ remain close to one.
For comparison, a subsystem constraint\cite{Verstichel2010,VanAggelen2011}
also adds one linear inequality per subsystem but requires auxiliary
variational calculations of the isolated subsystem for several electron
numbers, which are geometry dependent for multi-atom subsystems; the cuts used
here require only the HF density and the region overlap matrices.

\section{Conclusions}\label{sec:conclusions}
The drastic violation of size consistency in the dissociation of diatomic
molecules found by Van Aggelen \emph{et al.}\cite{VanAggelen2009,VanAggelen2010}
is largely removed, for the molecules and the basis sets studied here, by a
small number of linear conditions that need neither auxiliary calculations of
the fragments nor a prescribed charge for each fragment. At the separated-atom
limit, with the $P$, $Q$, $G$ conditions and the region cuts, the energy is
additive over the fragments to within \numSCOtherPqgReg~m$E_\mathrm{h}$ in
\numSCRestored{} of \numSCPoints{} systems.
For N$_2^+$ and CO$^+$, and for CO without the $T_1$ and $T_2$ conditions, a
nonadditive error remains; it may be related to the description of
spin-coupled fragments,\cite{Nakata2008,VanAggelen2012,LiDePrince2019} but its
cause has not been established here. The subsystem
constraints\cite{Verstichel2010,VanAggelen2011} achieve a similar correction
but require variational calculations of every subsystem for several electron
numbers, which limits their use in realistic applications.

Neither ingredient of the conditions studied here is new. The inequality
$(n-k)(n-k-1)\ge0$ for integers is a standard device in the theory of
$N$-representability,\cite{Erdahl1987} and the partition of a molecule into
atomic regions is the basis of the quantum theory of atoms in molecules, whose
zero-flux surfaces pass through the bond critical points, and of
electron-number distribution functions.\cite{Bader1991,Fradera1999,Francisco2007}
The half spaces used here are bounded by the plane through the bond critical
point and coincide with the zero-flux surface only approximately; this does
not affect the validity of the cuts, which holds for any sharp region. Their combination gives linear conditions on the
2-RDM that are independent of the Hamiltonian, are constructed from the HF
density alone, and address directly the electron-number distribution of a
fragment, which the $P$, $Q$, $G$, $T_1$, and $T_2$ conditions leave
insufficiently constrained in dissociation.

The calculations distinguish two ways in which the relaxation fails. In
heteronuclear ions the mean population of a fragment is wrong; the $T_1$ and
$T_2$ conditions do not change this, whereas a region cut at the bond critical
point restores an essentially integer fragment charge and removes most of the
energy error. In the homonuclear cation N$_2^+$ the mean population is correct
but its fluctuation is suppressed; the region cut restores the fluctuation,
and the error that remains is largely removed by the $T_1$ and $T_2$ conditions. In
both cases at most \numMaxRegionCuts{} region cuts were violated and added at
any geometry. How effective a cut is depends on the partition and on the
basis; the present data do not single out one partition as generally
preferable. In the minimal-basis systems studied, the
effective partitions, the bond-critical-point region and orthogonalized atomic
orbitals, are those for which the condition is nearly saturated by the exact
state. Orbitals localized separately in the occupied and virtual HF spaces may
not represent a fragment (N$_2^+$) or may leave a slack at the exact state
(NO$^+$, CN$^-$ in the minimal basis), and the resulting cuts can then be
satisfied without correcting the mean population; in the split-valence basis,
however, the orbital cuts of NO$^+$ are nearly as effective as the region cut at
long bond lengths. Where the localized HF orbitals give a nearly
saturated condition, they are also more sensitive than the half-space region to
the small charge transfer in neutral CO at intermediate bond lengths.

These conclusions rest on diatomic molecules in a minimal basis, three of
which (BeB$^+$, NO$^+$, and LiF) were also treated in a split-valence basis, and on
one lattice model, for which the cuts have essentially no effect. In the
split-valence basis the region cuts remain effective for BeB$^+$ and NO$^+$, and
neither partition changes the energy of LiF appreciably, because its relaxation
already follows the ionic--covalent crossing; compared with the
subsystem constraints for BeB$^+$, they give smaller errors at intermediate
bond lengths and larger ones at long bond lengths. Small polyatomic
systems in the minimal basis, and the generation of conditions on the joint
distribution of several regions by separation instead of enumeration, are
treated in Part II; larger basis sets with polarization functions are left for
future work. Extension to
larger molecules, including transition-metal systems, is in progress.

\section*{Supplementary Material}
The supplementary material contains the energy errors and the primal--dual
gaps at all computed points (Table~S1), including the scans of N$_2^+$ and
CO$^+$ with the HF occupations of the $^2\Sigma_g^+$ and $^2\Pi$
configurations; a check of the region overlap matrices and of the region-cut
energies against the integration grid (Table~S2); intervals for the
size-consistency errors (Table~S3); the accuracy of representative SDP
solutions (Table~S4); the fragment charges and dipole moments (Tables~S5 and
S6); the expression used for the local spin of a region; the energy
errors of BeB$^+$ and NO$^+$ in the split-valence basis at all points
(Table~S7); the composition of the orbital candidate sets (Table~S8); and the
energies with the region cut against the position of the plane (Table~S9);
the charge of the Li region in LiF along the STO-6G scan; and the energy
errors of LiF in the split-valence basis (Table~S10).

\begin{acknowledgments}
Some of the calculations were run on the HOKUSAI BigWaterfall2
supercomputer at RIKEN.
\end{acknowledgments}

\section*{Author Declarations}
\subsection*{Conflict of Interest}
The author has no conflicts to disclose.

\subsection*{Use of AI Tools}
The author used Claude Code (Anthropic), an AI coding assistant, in this work:
to write and run the scripts that generate the semidefinite-programming
inputs, collect the results, and produce the tables and figures; to search for
and check bibliographic data; and to draft parts of the manuscript text. The
author also used OpenAI's GPT-6 Astra model, through ChatGPT, mainly to review
drafts of the manuscript: its comments on the arguments, the numerical evidence, and the
literature were checked by the author against the data and the cited sources
before any change was made. All
numbers in the manuscript are generated by scripts in the accompanying
repository and are checked automatically against the stored data. The author
designed the study, directed and checked this work, and takes full
responsibility for the content of this article.

\section*{Data Availability}\label{rev:data}
The data and scripts that support the findings of this study are openly
available in Zenodo at \url{https://doi.org/10.5281/zenodo.24573421}, reference
number 10.5281/zenodo.24573421.\cite{NakataZenodoPartI} They
include all data needed to reproduce every number, table, and figure, in the
\texttt{data/} and \texttt{scripts/} directories; the frozen inputs from which
every SDPA input file is rebuilt and checked; and the inputs and per-point
results of the split-valence calculations.

\bibliography{references}
\end{document}